\documentclass[
    reprint,
    superscriptaddress,
    amsmath,
    amssymb,
    aps,
    pre,
    floatfix
]{revtex4-2}

\usepackage{graphicx}
\usepackage{bm}
\usepackage{dcolumn}
\usepackage{xcolor}
\usepackage[caption=false]{subfig}
\usepackage{hyperref}

\begin{document}

\title{
Infection Spreading and Stochastic Extinction in Finite Spatially Inhomogeneous Susceptible--Infected--Susceptible and Susceptible--Infected--Recovered Systems
}

\author{Sara Shabani}
\email{sarashabani@vt.edu}
\affiliation{
Department of Physics, 
Center for Soft Matter and Biological Physics,
Center for the Mathematics of Biosystems,
Virginia Tech, Blacksburg, Virginia 24061, USA
}

\author{Uwe C. T{\"a}uber}
\email{tauber@vt.edu}
\affiliation{
Department of Physics, 
Center for Soft Matter and Biological Physics,
Center for the Mathematics of Biosystems,
Virginia Tech, Blacksburg, Virginia 24061, USA
}
\affiliation{
Faculty of Health Sciences, Virginia Tech,
Blacksburg, Virginia 24061, USA
}
\date{\today}

\begin{abstract}
Demographic fluctuations can drive finite epidemic populations into an absorbing disease-free state even when the corresponding infinite system supports sustained activity. 
By means of Monte Carlo simulations, we investigate how spatial coupling of two distinct subsystems modifies stochastic extinction and epidemic spreading in susceptible–infectious–susceptible (SIS) and susceptible–infectious–recovered (SIR) models on inhomogeneous two-dimensional square lattices. 
For isolated SIS systems, extinction times exhibit approximate power-law scaling near the epidemic threshold even at comparatively small system sizes. 
System traversal times further characterize the distinct propagation dynamics of SIS and SIR infection clusters. 
We next consider diffusively coupled regions with different transmission characteristics in coupled SIS–SIS, SIR–SIS, and SIR–SIR configurations. 
Invading epidemic fronts can reactivate a vulnerable SIS region after its initial infection has disappeared and establish a long-lived active epidemic state. 
This stabilization can be induced by either a sustained SIS source or a single transient SIR epidemic wave. 
Spatial correlation measurements support a mechanism in which the invading activity generates an extended correlated infectious population, effectively providing favorable initial conditions for renewed epidemic spreading.
Correlations develop particularly strongly parallel to the subsystem interfaces. 
Decoupling the regions after some elapsed period demonstrates that autonomous persistence requires sufficient coupling time beyond the first interface arrival for this correlated population to become established. 
Increasing system size delays invasion but enhances the probability of ultimate successful stabilization. 
In two coupled SIR regions, transmission can initiate a macroscopic secondary outbreak, while permanent immunity restricts both epidemics to transient activity. 
Our results identify the spatial structure of invading infection clusters, the coupling duration, and the receiving region’s dynamics as central factors governing epidemic reactivation and long-lived persistence in finite heterogeneous populations subject to stochastic extinction events.
\end{abstract}

\maketitle
\section{Introduction}

A finite stochastic system with an absorbing state will eventually reach it, whence its subsequent time evolution terminates \cite{vanKampen1981}.
Specifically, this applies to stochastic epidemic dynamics in finite populations that are inevitably subject to extinction events driven by demographic fluctuations. 
Even when control parameters are chosen such that the corresponding infinite system sustains persistent activity, finite realizations may reach absorbing configurations in which the infection has vanished. 
This is at variance with standard mean-field approximations that neglect stochastic fluctuations, demographic noise, and spatio-temporal correlations.
The ensuing (coupled) rate equations invariably mistake complete extinction states as unstable fixed points.
Moreover, their representation of a population through a continuous density overlooks its fundamental discrete nature that invalidates linear expansions near absorbing states.
Yet understanding how such extinction events may be delayed or prevented is of fundamental relevance for nonequilibrium statistical physics, (bio-)chemistry, ecology, and epidemiology, and consequently requires a fully stochastic model representation \cite{KeelingRohani2008,Tauber2014,Tauber2019, DistefanoShabaniTauber2025}.

In spatially extended systems, infection spreading proceeds through the growth and propagation of infectious clusters.
Near epidemic thresholds, these clusters exhibit large fluctuations and critical scaling behavior governed by absorbing-state universality classes.
The stochastic lattice SIS model is equivalent to the contact process originally introduced by Harris \cite{Harris1974}. It represents the simplest stochastic epidemic process with reinfection and undergoes a continuous active-to-absorbing phase transition belonging to the directed percolation (DP) universality class \cite{Janssen1981,Grassberger1982}. The robustness of the DP universality class has also been demonstrated experimentally in transitions between turbulent and absorbing states in electrohydrodynamic liquid-crystal systems \cite{TakeuchiEtAl2007,TakeuchiEtAl2009}. The SIR model, originally introduced by Kermack and McKendrick \cite{KermackMcKendrick1927} and often referred to as the general epidemic process, is instead associated with dynamic isotropic percolation (DIP) \cite{Grassberger1983,CardyGrassberger1985,TomeZiff2010,MukhamadiarovTauber2022}. 
These universality classes determine the scaling properties of epidemic spreading, survival probabilities, correlation lengths, and characteristic time scales near their epidemic thresholds \cite{Hinrichsen2000,Odor2004, JanssenTauber2005,HenkelHinrichsenLubeck2008,Tauber2014}.

While critical behavior in homogeneous epidemic systems has been extensively studied, considerably less attention has been devoted to spatially inhomogeneous environments and interface phenomena in nonequilibrium dynamics featuring absorbing states \cite{SerraoTauber2021,DistefanoShabaniTauber2025}. 
Realistic populations are rarely homogeneous and often consist of regions with distinct transmission characteristics or environmental conditions and hence population densities. 
Understanding how epidemic activity propagates across interfaces separating regions with different dynamical properties therefore represents an important open problem in both statistical physics and epidemiology.

In this work, we investigate stochastic extinction and stabilization mechanisms in finite spatially extended SIS and SIR systems. 
We focus on systems composed of dynamically distinct regions that are diffusively coupled through local infection spreading across their interfaces. 
One subsystem is chosen close to the epidemic threshold and is therefore vulnerable to stochastic extinction, while the neighboring subsystem resides deeper in the active phase and effectively acts as a source of epidemic activity. 
We study how this dynamical coupling between spatially separated regions modifies extinction probabilities, extinction times, and the long-time persistence and intensity of epidemic activity.

Near the epidemic threshold, the correlation length $\xi$ becomes large and may approach the system size $L$, $\xi \sim L$. 
In this regime, fluctuations generated in the active subsystem can influence neighboring vulnerable regions over macroscopic distances. 
The corresponding relaxation dynamics is further drastically decelerated by critical slowing-down, characterized by the algebraic growth of the characteristic relaxation time $\tau \sim \xi^{z}$, where $z$ denotes the dynamic critical exponent. 
As a consequence, spatial coupling may substantially alter the fate of finite systems that would otherwise undergo stochastic extinction. 
Our goal is to determine under which conditions activity generated in a stable region can prevent extinction in a neighboring vulnerable subsystem and to identify the mechanisms responsible for this stabilization.

\section{Models and Methods}

\subsection{Stochastic lattice epidemic models}

We consider stochastic epidemic dynamics defined on two-dimensional square lattices of linear size $L$ with periodic boundary conditions. 
Each lattice site $i$ hosts one individual (local carrying capacity $K=1$), corresponding to a susceptible ($S$), infectious ($I$), or recovered ($R$) state, depending on the model.

In the susceptible--infectious--susceptible (SIS) model, the stochastic reactions are
\begin{align}
S + I &\xrightarrow{\lambda} I + I \, , \\
I &\xrightarrow{\gamma} S \, ,
\end{align}
where the first reaction describes the infection processes, restricted to nearest-neighbor lattice sites. 
$\lambda$ denotes the infection probability and $\gamma$ the recovery probability. 
The total population number is conserved, and the absorbing state corresponds to disease extinction ($n_I = 0$). 
This active-to-absorbing phase transition at the epidemic threshold is governed by the universal scaling exponents of critical directed percolation (DP) \cite{Janssen1981,Grassberger1982,Hinrichsen2000,Odor2004}.

For the susceptible--infectious--recovered (SIR) model, the recovery process (2) is replaced with
\begin{equation}
I \xrightarrow{\gamma} R \, ,
\end{equation}
which permanently removes recovered and immune individuals from the epidemic dynamics. 
Consequently, the SIR model sustains at most a single spreading epidemic wave and ultimately reaches an absorbing configuration devoid of infectious individuals. 
The critical dynamics near the SIR epidemic threshold is described by the dynamic isotropic percolation (DIP) universality class \cite{Grassberger1983,CardyGrassberger1985,TomeZiff2010,MukhamadiarovTauber2022}.

\subsection{Spatial coupling and inhomogeneous systems}

To investigate how spatial heterogeneity influences epidemic persistence and stochastic extinction, we construct spatially inhomogeneous systems composed of two diffusively coupled subsystems. 
Each subsystem occupies a rectangular region of the (toroidal) lattice and may either follow the same or different stochastic infection and recovery rules, but will definitely be characterized by different control parameters, most notably the infection probability $\lambda$.

While the subsystems are coupled, periodic boundary conditions apply within each subsystem in the direction parallel to the interfaces.
In the perpendicular direction, the right edge of each subsystem is connected to the left edge of the other, thus forming two interfaces. 
Once an infected individual enters the adjacent subsystem, it immediately obeys the local reaction processes and probabilities implemented there. 
In selected simulations, the diffusive coupling between the two subsystems is removed at a prescribed decoupling time $t_{\mathrm{dec}}$. After decoupling, each subsystem evolves independently with periodic boundary conditions implemented in both directions.
 
This framework enables us to investigate how a dynamically stable subsystem influences the spatio-temporal spreading and persistence of epidemic activity in a neighboring subsystem that is vulnerable to stochastic extinction. 
Of particular interest is the regime near the epidemic threshold, where stochastic fluctuations may drive small systems toward the inactive state, yet large correlation lengths and critical fluctuations may enhance the influence of one subsystem on the other, potentially inducing mutual stabilization.
The decoupling procedure allows us to determine whether epidemic activity induced in the vulnerable subsystem persists autonomously after transmission from the neighboring subsystem ceases.

\subsection{Monte Carlo dynamics and observables}

The stochastic dynamics is implemented using random sequential Monte Carlo updates. 
During each elementary update, a lattice site is selected uniformly at random and the corresponding infection or recovery reaction is applied probabilistically according to the state of the particle and its nearest neighbors.

We monitor the time-dependent densities
\begin{equation}
n_X(t) = \frac{1}{L^2} \sum_i \delta_{s_i,X}, \qquad
X \in {S,I,R},
\end{equation}
as well as the survival probability $P_{\mathrm{surv}}(t)$, defined as the large-ensemble average of the fraction of realizations that have not yet reached an absorbing state by time $t$.
Throughout this paper, time is measured in Monte Carlo steps (MCS).
For simulations performed near the epidemic threshold, we additionally analyze extinction times, extinction probabilities, and the spatial propagation of epidemic activity. 
In coupled systems, we further characterize the transmission of activity across subsystem interfaces through measurements of system traversal times and the subsequent evolution of the coupled dynamics \cite{SwailemEtAl2025}.

----------------------------------------------------------
\subsection{Initial conditions and simulation protocols}

The stochastic dynamics exhibits a pronounced sensitivity to the chosen initial conditions, particularly in finite systems close to absorbing states such as epidemic thresholds. 
To systematically assess this dependence, we consider several distinct sets of initial configurations.
Unless stated otherwise, simulations are initialized with all lattice sites occupied by susceptible individuals, except for a single (or a small number of) infectious seed(s) placed near the center of each subsystem.
To investigate the role of correlated initial states, we also consider spatially extended initial conditions in which a finite fraction of sites is randomly assigned to the infectious state (according to a Poisson distribution). 
These configurations mimic scenarios with more widespread initial exposure that reduce the likelihood of early stochastic extinction.

In spatially inhomogeneous systems, each subsystem is initialized independently according to its local reaction rate parameters. 
This allows us to disentangle extinction events driven by unfavorable initial conditions from those caused by intrinsic stochastic instability.
All simulations are evolved using identical Monte Carlo steps, and observables are averaged over large ensembles of $500$ independent realizations to ensure statistical reliability.

\section{Infectious Kinetics in Homogeneous Systems: Quantitative Characterization}

We first consider isolated SIS systems initialized with localized infectious seeds. 
Even for parameter values that support sustained activity in the thermodynamic limit, sufficiently small finite systems frequently undergo extinction due to stochastic fluctuations that drive them towards the absorbing state.
These fluctuations incorporate demographic noise:
In contrast with the mean-field rate equation description in terms of continuous species densities, for which the total extinction state is invariably unstable, chemical reactions, population dynamics, and epidemic kinetics entail integer population numbers.
This fundamental discreteness renders the absorbing configuration stable and in fact, the only stationary state in the stochastic long-time evolution \cite{vanKampen1981,DobrinevskiFrey2012,Tauber2014,DistefanoShabaniTauber2025}. 

Away from the critical point and inside the active phase, the mean extinction time may increase exponentially with the system volume. Thus,
although extinction remains inevitable in a finite stochastic system in the infinite-time limit, the active state may constitute an exceedingly long-lived quasi-stationary state \cite{Tauber2014,DistefanoShabaniTauber2025}.
Near the epidemic threshold, however, the divergence of correlation lengths and relaxation times becomes algebraic and is governed by universal absorbing-state critical scaling.

In the SIS model close to the epidemic threshold, realizations of the stochastic kinetics exhibit strong variability. 
While some trajectories develop expanding infection fronts and reach long-lived active states, others are driven into the absorbing disease-free configuration. 
The probability of extinction increases markedly upon reducing the system size or tuning the infection probability closer to its critical value $\lambda_c$.
In this section, we investigate how the extinction time and fraction of extinct realizations of an isolated, spatially homogeneous system depend on the distance from the epidemic threshold. 
Remarkably, we observe critical power laws already for comparatively small system sizes.
In addition, we explore the initial growth law for the epidemic and its characteristic spreading front speed.

\subsection{Extinction time: near-threshold  scaling}

Close to the critical infection probability $\lambda_c$, the stochastic spatially extended epidemic dynamics is governed by universal scaling laws characteristic of the directed percolation (DP) universality class \cite{Hinrichsen2000,Odor2004,Tauber2014}. 
In an infinite system, the characteristic relaxation time scale diverges as a power law when approaching the critical point,
\begin{equation}
\tau \sim \xi^z \sim |\lambda - \lambda_c|^{-\nu_\parallel} \, ,
\end{equation}
where $\nu_\parallel = z \nu_\perp$ denotes the temporal correlation exponent, where the correlation length diverges according to $\xi \sim |\lambda - \lambda_c|^{-\nu_\perp}$. 
This drastic critical slowing-down reflects the emergence of long-lived fluctuations and persistent correlations at the continuous phase transition separating the active and absorbing state, and indicates the absence of an intrinsic (microscopic) characteristic time scale at $\lambda_c$.
For two-dimensional spatial systems, the accepted DP estimates for these critical exponents are $\nu_\perp \approx 0.733$, $\nu_\parallel \approx 1.295$, and $z = \nu_\parallel / \nu_\perp \approx 1.766$ \cite{Hinrichsen2000,Odor2004}.

\begin{figure}[t]
  \centering
  \includegraphics[width=0.95\columnwidth]{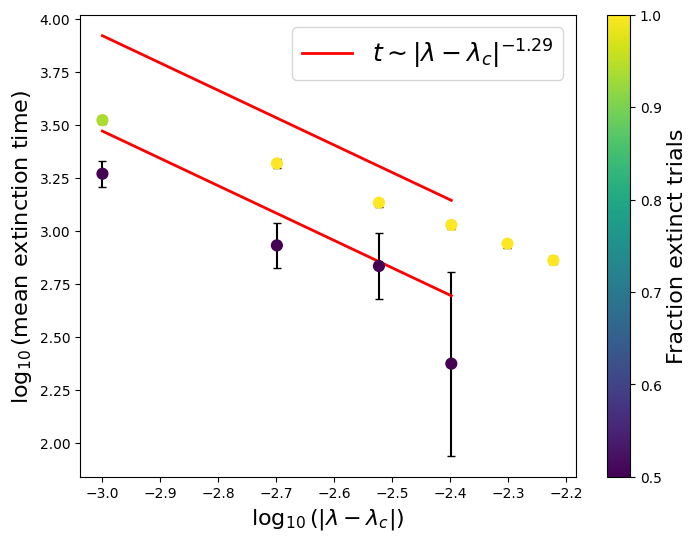}
  \caption{
    Mean extinction time $\tau$ as a function of the distance from the epidemic threshold $|\lambda - \lambda_c|$ for isolated SIS systems with fixed linear system size $L=100$ and initial infected density $\rho_0 = I_0/L^2 = 0.01$. 
    The double-logarithmic data representation reveals approximate power law scaling, consistent with the directed percolation (DP) prediction $\tau \sim |\lambda - \lambda_c|^{-\nu_\parallel}$ near criticality with $\nu_\parallel \approx 1.29$ in two dimensions (red lines). 
    The error bars represent statistical uncertainties obtained from $500$ independent realizations. 
    The color scale indicates the fraction of realizations that reached the absorbing state within $10,000$ MCS.
    The darker (black/blue) markings correspond to data obtained in the active epidemic state ($\lambda>\lambda_c$), whereas the lighter (yellow/green) points were taken on the inactive, absorbing side of the transition ($\lambda < \lambda_c$).
    }
  \label{fig:extinction_lambda_scaling}
\end{figure}
For each infection probability value $\lambda$, we performed $500$ independent Monte Carlo runs to sample over different stochastic realizations and recorded the first passage time to reach the absorbing state.
Figure~\ref{fig:extinction_lambda_scaling} shows the resulting mean extinction time $\tau$ measured in a square SIS system with linear size $L=100$ as a function of the distance $|\lambda - \lambda_c|$ from the critical infection probability $\lambda_c$. 
The initial infected population density was set at $\rho = I_0/L^2 = 0.01$. 
Representing the measured mean extinction times $\tau$ in a double-logarithmic plot reveals clear power-law scaling behavior over a broad range of infection parameters $\lambda$ already in this fairly small system. 
A linear fit to the data in this graph yields an exponent consistent with the prediction $\nu_\parallel \approx 1.29$ on both the active- and absorbing-state sides of the phase transition, {\em i.e.}, in excellent agreement with the known value for two-dimensional directed percolation.
Note that the color scale in Fig.~\ref{fig:extinction_lambda_scaling} indicates the fraction of realizations that reached extinction during the observation window, extending to $10,000$ MCS.

Physically, this power law behavior arises because in a sufficiently large lattice situated near the epidemic threshold, infected clusters grow and decay over increasingly long time scales. 
As $\lambda$ approaches $\lambda_c$ from below (appearing in lighter yellow--green shades), epidemic activity becomes marginally stable, and stochastic fluctuations require progressively longer times to drive the system into the absorbing state.
For infection probabilities farther away from the critical point, extinction occurs more frequently and more rapidly, as epidemic spreading is less strongly constrained by critical fluctuations. 
Conversely, closer to the threshold, large correlated regions and slowly propagating infection fronts significantly prolong the survival time.

These results confirm that extinction dynamics in comparatively small finite SIS systems near the epidemic threshold is governed by universal directed percolation scaling, with the mean extinction time displaying critical slowing-down, {\em i.e.}, diverging algebraically as the control parameter approaches its critical value.

\subsection{Extinction time: critical finite-size scaling}

To quantify the dependence of stochastic extinction on linear system size $L$, we systematically measured the mean extinction time $\tau(L)$ for isolated SIS systems at fixed infection probability close to the epidemic threshold $\lambda_c = 0.106$. 
For each $L$, we performed $500$ independent stochastic realizations and recorded the mean first-passage time $\tau$ to the absorbing state.

\begin{figure}[t]
  \centering
  \includegraphics[width=0.95\columnwidth]{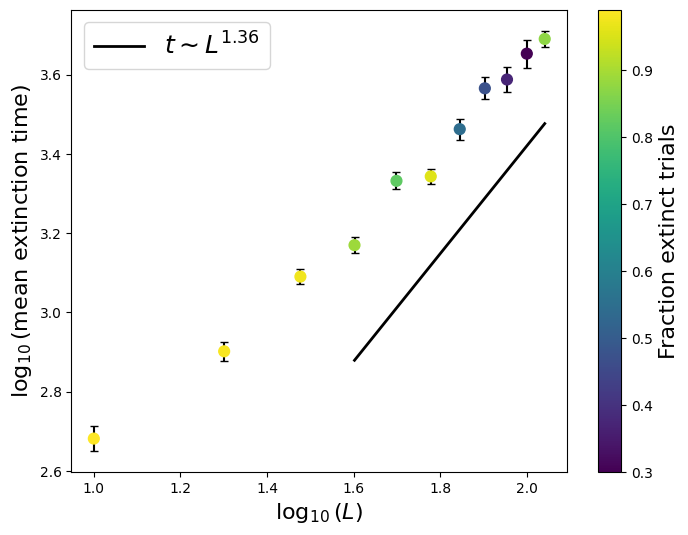}
  \caption{
    Mean extinction time $\tau$ as a function of linear system size $L$ for isolated SIS systems with the infection probability fixed at the (quasi-)critical value $\lambda = \lambda_c(L)$ and initial infected density $\rho_0 = I_0/L^2 = 0.01$. 
    The error bars represent statistical uncertainties from $500$ independent realizations. 
    A fit to the expected power law $\tau(L) \sim L^{z}$ yields $z \approx 1.36$ for the dynamic critical exponent, deviating for these comparatively  small systems from the expecetd two-dimensional DP value $z = 1.766$.
    The color scale indicates the fraction of realizations reaching the absorbing state within $10,000$ MCS.
    }
  \label{fig:extinction_scaling}
\end{figure}
Figure~\ref{fig:extinction_scaling} shows the resulting mean extinction times plotted as a function of linear system size on double-logarithmic scale. 
The data exhibit the expected power law dependence,
\begin{equation}
\tau(L) \sim L^{z} \ ,
\end{equation}
where $z$ denotes the dynamic critical exponent governing temporal scaling at the absorbing-state phase transition.
A linear fit in the log-log representation yields the effective exponent $z_{\mathrm{eff}} \approx 1.36$. 
This value is smaller than the accepted two-dimensional DP exponent $z\approx1.766$ \cite{Hinrichsen2000,Odor2004}, indicating the presence of appreciable finite-size, finite-time, or off-critical corrections over the presently accessible parameter range.
Yet this power-law behavior again reflects the appearance of critical slowing-down and the absence of an intrinsic characteristic time scale at criticality as $L \to \infty$. 
Instead, in finite lattices the temporal evolution is controlled by the system size $L$, which sets the maximum spatial extent of spanning infection clusters and consequently posits an upper limit for the lifetime of epidemic activity.

For small $L$, extinction occurs rapidly due to large relative population fluctuations. 
As the system size increases, spatial correlations and propagating infection fronts stabilize activity over increasingly longer time scales, resulting in the observed algebraic growth of the mean extinction time. 
The algebraic scaling behavior confirms that stochastic extinction in finite SIS systems near the epidemic threshold is governed by large-scale critical dynamics rather than microscopic model details.
Again, the color scale in Fig.~\ref{fig:extinction_scaling} indicates the fraction of realizations that reached extinction within the observation time window. 
As expected, extinction becomes less frequent with increasing system size, reflecting the enhanced stability of larger spatial systems against stochastic fluctuations.
Intriguingly, we observe the algebraic dependence $\tau \sim L^z$ at $\lambda_c(L)$ already for quite small systems; further away from the extinction threshold, one would instead typically expect exponential behavior $\log \tau \sim L^d$ (in $d$ spatial dimensions).

\subsection{Extinction time: finite-size scaling collapse}

The power-law scaling discussed in the previous two subsections can be unified within the framework of finite-size scaling theory. 
Near a continuous active-to-absorbing-state phase transition, the dependence of the characteristic extinction time on both the distance $|\lambda-\lambda_c|$ from the critical point and the finite system size $L$ is expected to obey a universal scaling relation. 
Rather than considering the infection probability and system size separately, finite-size scaling combines both quantities into a single scaling form.
For systems belonging to the directed percolation universality class, the mean extinction time is expected to satisfy
\begin{equation}
\tau(\lambda,L) = L^z \,
\hat{\tau}\big( |\lambda-\lambda_c| \, L^{1/\nu_\perp}\big) \, ,
\label{eq:FSS_tau}
\end{equation}
where $z$ and $\nu_\perp$ respectively denote the dynamic critical and spatial correlation length exponents, and $\hat{\tau}$ indicates a universal scaling function. 
This scaling form reflects the fact that sufficiently close to the epidemic threshold the correlation length is limited by the finite system size, such that the relevant scaling variable becomes the dimensionless combination $y=|\lambda-\lambda_c| \, L^{1/\nu_\perp}$.

\begin{figure}[t]
    \centering
    \includegraphics[width=0.95\columnwidth]{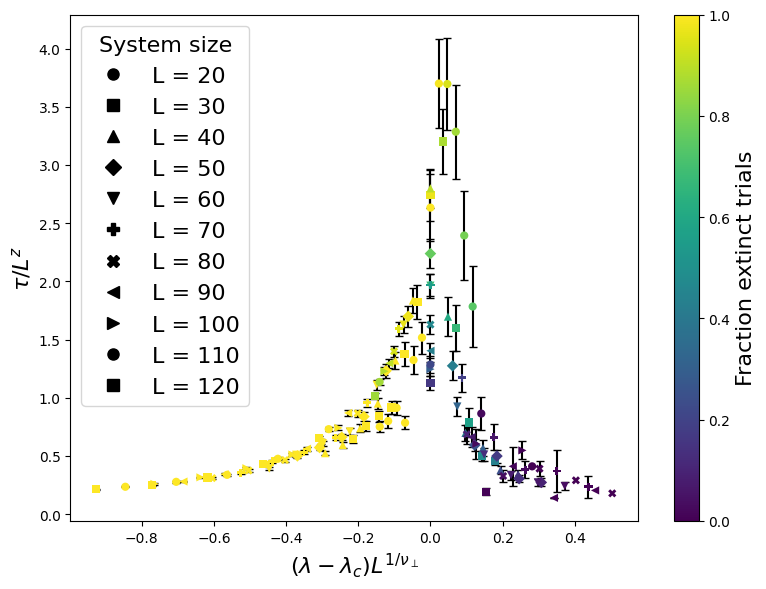}
    \caption{
    Finite-size scaling collapse of the mean extinction time for isolated SIS systems. 
    The mean extinction time is rescaled as $\tau/L^z$ and plotted as a function of the scaling variable $y = |\lambda-\lambda_c| \, L^{1/\nu_\perp}$ using the effective critical exponents extracted previously, $z \approx 1.36$ and $\nu_\perp = \nu_\parallel / z \approx 0.95$. 
    Different marker shapes denote different system sizes, while the color scale indicates the fraction of realizations that reached the absorbing state within the observation window.
    }
    \label{fig:fss_collapse}
\end{figure}
To test this prediction, we rescale the extinction times measured for different system sizes according to Eq.~(\ref{eq:FSS_tau}) using the effective critical exponents obtained from the previous two sections. 
From the finite-size scaling of the extinction time at the critical point, we obtained the dynamic exponent $z \approx 1.36$.
Similarly, fitting the extinction time as a function of the distance from the critical point yielded the temporal correlation exponent $\nu_\parallel \approx 1.29$.
Using the scaling relation $\nu_\parallel = z \nu_\perp$, we then estimate the corresponding spatial correlation-length exponent as $\nu_\perp = \nu_\parallel / z \approx 0.95$.
These numerically determined exponents are then used to rescale the extinction times for all simulated system sizes and infection probabilities.

The resulting scaling collapse is shown in Fig.~\ref{fig:fss_collapse}. 
Data obtained for different system sizes exhibit a reasonable collapse over most of the investigated parameter range, indicating that the extinction dynamics is well described by the finite-size scaling ansatz (\ref{eq:FSS_tau}). 
Although systematic deviations remain for the smallest system sizes, where
corrections to scaling are expected to be most pronounced, the data for
$L \gtrsim 50$ display a reasonable collapse. In this size range, the
extinction time is therefore primarily governed by the combined scaling
variable $|\lambda-\lambda_c|L^{1/\nu_\perp}$ rather than by the infection
probability and system size independently.
The scaling function displays two distinct branches corresponding to the active and absorbing sides of the epidemic threshold. 
As the critical point is approached from either side, the rescaled extinction time increases due to critical slowing-down and the growth of long-range spatio-temporal correlations. 
Away from the transition, the extinction time decreases because the system generally evolves on progressively shorter characteristic time scales.

Together with the power-law behavior observed in
Figs.~\ref{fig:extinction_lambda_scaling} and
\ref{fig:extinction_scaling}, the finite-size scaling collapse supports the
interpretation that the SIS extinction dynamics is governed by critical
scaling. Appreciable corrections remain for the smaller lattices, while the
collapse becomes considerably more consistent for $L\gtrsim50$.
However, the effective exponents employed here, $z_{\mathrm{eff}}\approx1.36$ and $\nu_{\perp,\mathrm{eff}}\approx0.94$, differ from the accepted values for directed percolation in two spatial dimensions, $\nu_\perp\approx0.733$, $\nu_\parallel\approx1.295$, and hence $z=\nu_\parallel/\nu_\perp\approx1.766$ \cite{Hinrichsen2000,Odor2004}.
The observed deviations are likely associated with the restricted range of accessible system sizes, the finite observation window, and corrections to asymptotic scaling. 
Nevertheless, the improved collapse for $L\gtrsim50$ supports the applicability of the finite-size scaling form, while the fitted exponents should be interpreted as effective finite-size estimates rather than asymptotic universal critical exponent values.
\subsection{Dynamic scaling of the correlation length}

The divergence of the extinction time near the epidemic threshold is accompanied by the growth of long-range spatial correlations. 
To characterize the spatial structure of epidemic spreading, we compute the equal-time spatial correlation function and extract the corresponding isotropic correlation length $\xi(t)$ from the averaged correlation function over $4000$ independent realizations. 
The resulting correlation length provides a direct measure of the characteristic size of correlated infection clusters as the epidemic evolves.

Dynamic scaling theory predicts that, at the critical point, the characteristic correlation length grows algebraically in time according to
\begin{equation}
\xi(t) \sim t^{1/z} \, ,
\end{equation}
where $z$ denotes the dynamic critical exponent of the directed percolation universality class. 
Consequently, a double-logarithmic representation of $\xi(t)$ as a function of time should yield a straight line with slope $1/z$.

\begin{figure}[t]
    \centering
    \includegraphics[width=0.95\columnwidth]{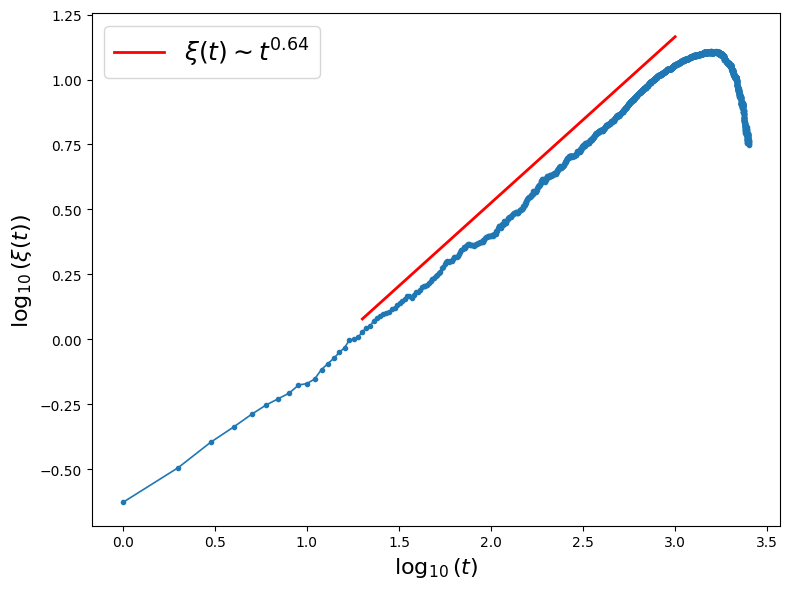}
    \caption{
    Dynamic growth of the isotropic correlation length $\xi(t)$ for an isolated SIS system at the epidemic threshold. 
    The correlation length is extracted from the averaged equal-time spatial correlation function obtained from $4000$ independent realizations. 
    The solid black line represents a power law fit $\xi(t) \sim t^{0.64}$, which yields an effective dynamic critical exponent $z \approx 1.56$.
    }
    \label{fig:xi_scaling}
\end{figure}
Figure~\ref{fig:xi_scaling} presents the time evolution of the isotropic correlation length on double-logarithmic scales. 
Following an initial transient regime, the correlation length exhibits an extended interval of algebraic growth before finite-size effects eventually limit its increase. 
A linear fit within the scaling regime yields $\xi(t)\sim t^{0.64}$, corresponding to an (effective) dynamic critical exponent $z \approx 1 / 0.64  \approx 1.56$.
Given the comparatively small system sizes explored here, this value is reasonably close to the effective dynamic exponent extracted independently from the finite-size scaling of the extinction time in Sec.~3.2, providing an additional consistency check of the dynamic scaling behavior. 
The slight discrepancy between the two estimates is expected since the correlation length is extracted from finite systems and finite observation times, where corrections to scaling and finite-size effects remain significant.

At later times, the correlation length deviates from the initial power law growth and eventually saturates. 
This crossover occurs when the growing correlation length becomes comparable to the finite system size $L$, preventing further development of long-range correlations. 
The subsequent fluctuations around a nearly constant value therefore reflect finite-size limitations rather than a breakdown of the underlying critical scaling.
\subsection{Extinction time: dependence on the initial infected density}

We next investigate the dependence of the mean extinction time on the initial number $I_0$ of infectious individuals. 
This dependence is particularly important near the epidemic threshold, where stochastic fluctuations compete with critical spreading dynamics.

\begin{figure}[t]
  \centering
  \includegraphics[width=0.95\columnwidth]{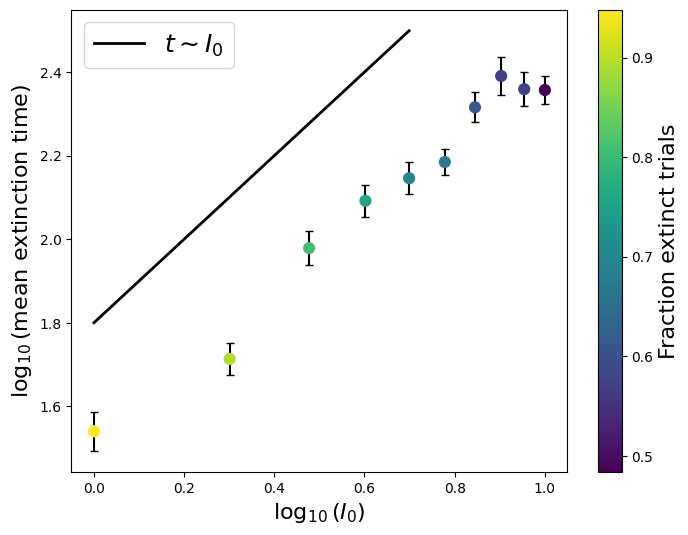}
  \caption{
    Mean extinction time $\tau$ as a function of the initial number of infected individuals $I_0 = n_I(t=0)$ for isolated SIS systems with fixed system size $L=100$ and infection probability set to the critical value $\lambda_c = 0.106$. 
    The double-logarithmic representation shows approximate power law scaling, indicating that the extinction time increases with the intensity of the initial infection. 
    Error bars represent statistical uncertainties from independent realizations. 
    The color scale indicates the fraction of realizations that reached the absorbing state within the observation window.
    }
  \label{fig:extinction_initial_scaling}
\end{figure}
Figure~\ref{fig:extinction_initial_scaling} shows the mean extinction time $\tau$ as a function of the initial number of infected individuals $I_0 = n_I(t=0)$ on a double-logarithmic scale for linear system size $L=100$ at the critical point $\lambda = \lambda_c$. 
The data exhibit clear power-law scaling $\tau \sim I_0^{\alpha}$ with an effective exponent $\alpha \approx 1$ over one to ten initially infected individuals that were randomly distributed on the lattice. 
For such small initial infected densities $\rho_0 = I_0/L^2 < 0.001$, this linear dependence can be understood in terms of independently spreading infection clusters originating from the few seeds that remain spatially separated at early times.
Consequently, their subsequent evolution proceeds approximately independently. 
In this dilute regime, increasing the initial number of infected individuals proportionally enhances the probability that at least one cluster survives for a longer duration, leading to an approximately linear increase in the mean extinction time. 

Moreover, a larger initial population also diminishes the relative importance of stochastic fluctuations, thereby stabilizing epidemic activity and delaying extinction.
This is evident in Fig.~\ref{fig:extinction_initial_scaling}, where the color scale again indicates the fraction of realizations that reached extinction within the observation time window.
As expected, for larger initial populations, extinction becomes less frequent, reflecting the increased robustness of epidemic spreading against stochastic fluctuations.



\subsection{Stochastic epidemic extinction in small systems}

Having established how extinction time and extinction probability depend on control parameters such as the infection probability, system size, and the number of initially infected individuals, we now focus on the stochastic extinction dynamics of a specific finite system that resides close to the absorbing epidemic threshold.

\subsubsection{Stochastic extinction dynamics in isolated SIS systems}
\label{sec:isolated_SIS_extinction}

To this end, we execute stochastic SIS dynamics on a square lattice of linear size $L=100$ with infection probability fixed at $\lambda = 0.109$, slightly above the critical value $\lambda_c(L=100)= 0.106$. 
Although this parameter choice situates the system in the active phase in the thermodynamic limit, a corresponding finite lattice remains susceptible to stochastic extinction due to demographic fluctuations. 
Hence, we refer to this setup as {\em vulnerable}.

\begin{figure}[t]
  \centering
  \includegraphics[width=0.95\columnwidth]{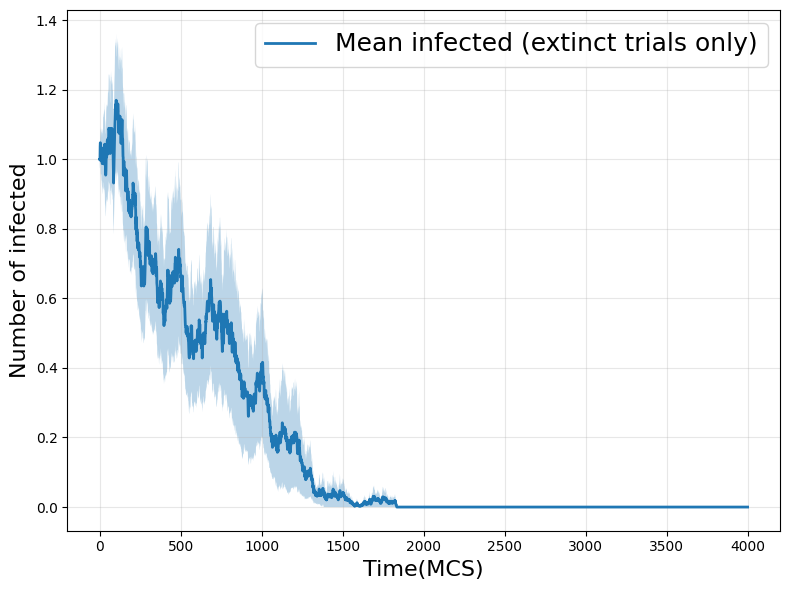}
  \caption{
    Time evolution of the mean number of infected individuals $\langle I(t)\rangle_{\mathrm{ext}}$ in an isolated SIS system of linear size $L=100$ at infection probability $\lambda=0.109$, slightly above the estimated thermodynamic-limit critical value $\lambda_c(\infty)\approx0.106$.
    The average is conditioned on realizations that reach the absorbing state during the observation window, and the shaded region indicates the standard error of the mean.
    The infected population decreases approximately linearly on average over the main transient regime, although substantial stochastic fluctuations and deviations from a simple functional form remain.
    Approximately $96\%$ of the realizations reach the absorbing state, with extinction generally occurring within $2,000$ MCS.
    }
  \label{fig:vulnerable_decay}
\end{figure}
Figure~\ref{fig:vulnerable_decay} shows the time evolution of the infectious population in this small isolated system. 
Averaging over realizations that reach extinction reveals a rapid decay of the infectious density $I(t) / L^2$, with extinction typically occurring by $t \approx 2,000$ MCS. 
In our simulations, approximately $96\%$ of realizations eventually reached this absorbing state, confirming that stochastic fluctuations dominate the dynamics in this parameter regime despite its parameters residing in the active phase, \textit{i.e.}, above the critical threshold $\lambda_c$, in the thermodynamic limit $L \to \infty$.
The decay of the infectious density reflects the progressive collapse of small active infected clusters driven by stochastic fluctuations. 
Initially, the infection may spread locally, but in the absence of sustained large-scale activity, fluctuations eventually eliminate all infected individuals. 
This behavior highlights the fundamental difference between finite and infinite systems: 
While epidemic activity is stable above the critical point in the thermodynamic limit, finite systems inevitably undergo extinction on sufficiently long time scales.

\subsubsection{Extinction criteria for isolated SIR systems}
\label{sec:isolated_SIR_extinction}

A direct repetition of the preceding SIS extinction analysis is not appropriate for the SIR model. 
In any finite SIR system, the infectious population ultimately vanishes because infected individuals recover irreversibly and cannot become susceptible again. 
Hence, $n_I=0$ is the expected final state of both an outbreak that terminates near its initial seed and a successful epidemic that spreads across a macroscopic fraction of the system. 
The long-time extinction probability is therefore equal to one and cannot provide a useful measure of epidemic success.

Instead, isolated SIR realizations must be classified according to the spatial extent and cumulative size of the transient epidemic cluster. 
Useful observables include the probability that the infection cluster reaches the system boundary, the final recovered fraction $\rho_R(\infty)$, and the time required for a boundary-reaching epidemic to traverse the system. 
Boundary arrival indicates that the epidemic cluster has spread over a distance of order $L$ and therefore provides a natural finite-size criterion for a macroscopically developed SIR outbreak.

In Sec.~\ref{sec:traversal_time}, we consequently compare isolated SIS and SIR dynamics through their system traversal times and boundary-arrival probabilities as functions of both infection probability and system size.
These quantities replace the long-time survival criterion used for SIS dynamics and allow failed, spatially localized SIR outbreaks to be distinguished from epidemic clusters that manage to span the entire system.

\subsection{Infection front propagation and traversal time}
\label{sec:traversal_time}

To quantify how quickly epidemic activity propagates spatially through an isolated system, we measure the mean time required for infection fronts originating from a central infectious seed to first reach the system boundary. 
We refer to this quantity as the \textit{system traversal time} $t_f$. 

The expected scaling of the traversal time depends on whether the spreading process is considered directly at the epidemic threshold or in the supercritical regime. 
For a ballistically spreading front with constant propagation speed $v_f$, the traversal time across a system with linear extension $L$, starting from its center, is approximately
\begin{equation}
t_f = L / 2v_f \ .
\label{eq:tf_velocity}
\end{equation}
Thus, sufficiently far inside the active phase, where a well-defined asymptotic front velocity exists, the traversal time grows linearly with $L$.

For the spreading of critical clusters at the epidemic threshold, however, there exists no finite asymptotic propagation velocity. 
Instead, the characteristic spatial extent of the spreading cluster grows according to
\begin{equation}
R(t) \sim t^{1/z} \ ,
\end{equation}
where $z$ denotes the dynamic critical exponent. 
Requiring the cluster to spread over a distance of order $L/2$ therefore yields
\begin{equation}
t_f(L) \sim L^z \ .
\label{eq:tf_L}
\end{equation}
Since $1<z<2$ for the critical spreading processes considered here, this corresponds to propagation that is considerably slower than ballistic motion, $t_f\sim L$, but still faster than ordinary diffusion, $t_f\sim L^2$, \textit{i.e.}, critical spreading is superdiffusive.

A complementary scaling behavior is expected when the system size is held fixed while the infection probability is varied above the epidemic threshold \cite{Hinrichsen2000,Tauber2014}. 
In this regime, the propagation can be characterized by a finite front velocity $v_f$ that vanishes upon approaching the critical point. 
The characteristic spatial and temporal correlation scales diverge as
\begin{equation}
\xi_\perp \sim \Delta^{-\nu_\perp}, \qquad
\xi_\parallel \sim \Delta^{-\nu_\parallel},
\end{equation}
where $\Delta \sim |\lambda - \lambda_c|$ denotes the distance from the corresponding epidemic threshold. 
Since the characteristic velocity scales as the ratio of these spatial and temporal scales,
\begin{equation}
v_f \sim \xi_\perp / \xi_\parallel \sim \Delta^{\nu_\parallel-\nu_\perp} \ ,
\label{eq:front_velocity_scaling}
\end{equation}
the traversal time at fixed $L$ is expected to scale as
\begin{equation}
t_f \sim L / v_f \sim \Delta^{-(\nu_\parallel-\nu_\perp)} \ .
\label{eq:tf_delta}
\end{equation}
The measurements of $t_f$ as a function of system size and distance from the epidemic threshold therefore probe two complementary aspects of epidemic spreading, namely the anomalous dynamic scaling of critical clusters and the emergence of a finite propagation velocity in the supercritical regime.

\subsubsection{SIS dynamics}

We first consider the SIS model, whose active-to-absorbing phase transition belongs to the directed percolation (DP) universality class. 
Starting from a single infectious seed at the center of the lattice, we record the first time at which epidemic activity reaches the system boundary. 
The mean traversal time is calculated from the realizations that successfully reach the boundary, while the fraction of realizations that fail to do so within the observation window is shown by the color scale.
\begin{figure*}[t]
    \centering
    \subfloat[
        Dependence of $t_f$ on $\lambda-\lambda_c$.
        \label{fig:tf_SIS_delta}
    ]{
        \includegraphics[width=0.47\textwidth]
        {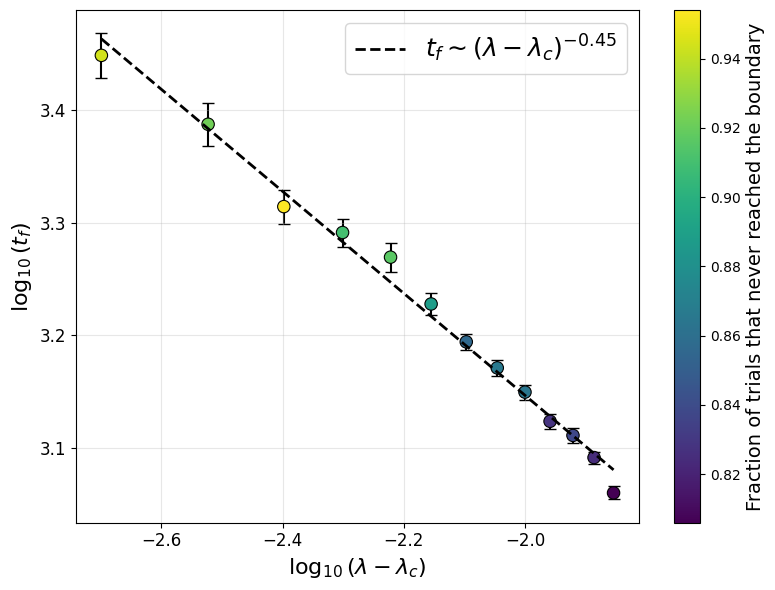}
    }
    \hfill
    \subfloat[
        Dependence of $t_f$ on system size $L$.
        \label{fig:tf_SIS_L}
    ]{
        \includegraphics[width=0.47\textwidth]
        {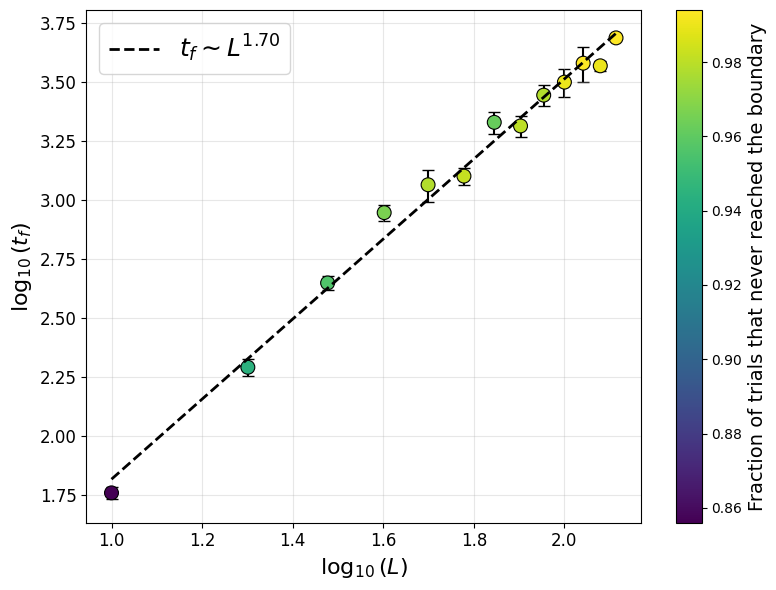}
    }
    \caption{
    Mean system traversal time $t_f$ for isolated SIS dynamics initialized with a single infected individual at the center of the lattice.
    (a) Double-logarithmic representation of $t_f$ as a function of the distance from the epidemic threshold, $\Delta = \lambda-\lambda_c$, for $\lambda > \lambda_c$, at fixed linear system size $L=100$.
    (b) Double-logarithmic representation of $t_f$ as a function of $L$ at the critical infection probability $\lambda=\lambda_c=0.106$.
    For each parameter value, the mean traversal time is calculated only from realizations in which the spreading infection reaches the system boundary.
    Error bars indicate the statistical uncertainties obtained from the successful realizations, while the color scale shows the fraction of  realizations in which the infection does not reach the boundary within the observation time interval.
    }
    \label{fig:tf_SIS}
\end{figure*}

Figure~\ref{fig:tf_SIS_delta} shows the traversal time as a function of the distance $\Delta=\lambda-\lambda_c$ from the SIS epidemic threshold. 
The traversal time decreases systematically as the infection probability is increased: 
Larger values of $\lambda$ facilitate the growth of infection clusters and accelerate the propagation of activity across the lattice, allowing the infection front to reach the system boundary more rapidly. 
Conversely, as $\lambda$ approaches $\lambda_c$ from above, propagation becomes increasingly slow and $t_f$ grows.
A linear fit in the double-logarithmic representation yields $t_f \sim (\lambda-\lambda_c)^{-0.453\pm0.013}$.
The observed increase in $t_f$ toward the transition is therefore associated with the progressive reduction of the supercritical propagation velocity rather than simply with the bulk relaxation-time divergence $|\lambda-\lambda_c|^{-\nu_\parallel}$. 
According to Eq.~(\ref{eq:tf_delta}), the asymptotic critical prediction for a propagating front instead involves the exponent difference $\nu_\parallel-\nu_\perp = (z - 1) \nu_\perp$.

The measured effective exponent is also numerically close to $1/2$, which is characteristic of the square-root dependence of the front velocity on the distance from threshold in mean-field reaction-diffusion descriptions \cite{Murray2002}. 
The present stochastic spatial system, however, is subject to strong fluctuations near the absorbing-state transition. 
The measured value should therefore be interpreted as an effective front-propagation exponent over the finite parameter range investigated here, with possible crossover between mean-field-like front propagation farther inside the active phase and fluctuation-dominated DP scaling sufficiently close to the critical point.

The comparatively long traversal times observed for the SIS model can also be understood from its microscopic dynamics. 
In SIS kinetics, infected individuals recover and immediately become susceptible again, allowing reinfections to occur repeatedly within the same spatial region. 
Consequently, infection clusters do not propagate strictly outward but instead undergo repeated local spreading, recovery, and reinfection processes. 
The resulting activity front is therefore less sharply directed than an epidemic wave that continuously advances into previously unaffected territory.

To independently probe the critical spreading dynamics, we next fix the infection probability at the epidemic threshold and vary the system size. 
Figure~\ref{fig:tf_SIS_L} shows that the corresponding traversal time increases algebraically with $L$. 
A linear fit in the double-logarithmic representation yields $t_f(L) \sim L^{1.696\pm0.050}$. 
The resulting effective exponent is close to the numerical estimate
$z=1.7665(2)$ for directed percolation in two spatial dimensions
\cite{WangEtAl2013}.
Thus, when measured directly through the spatial propagation of critical infection clusters, the system traversal time displays the expected anomalous dynamic scaling $t_f \sim L^z$.
The value obtained from the traversal-time scaling is also closer to the accepted DP exponent than the effective exponent extracted earlier from the finite-size dependence of the extinction time. 
This difference indicates that the traversal time provides a particularly direct measure of the relation between spatial and temporal scales during critical spreading, whereas extinction times in finite systems may be more strongly affected by finite-size corrections, the initial condition, and the finite observation time window.

\begin{figure*}[t]
    \centering
    \subfloat[
        Dependence of $t_f$ on $\lambda-\lambda_c$.
        \label{fig:tf_SIR_delta}
    ]{
        \includegraphics[width=0.47\textwidth]
        {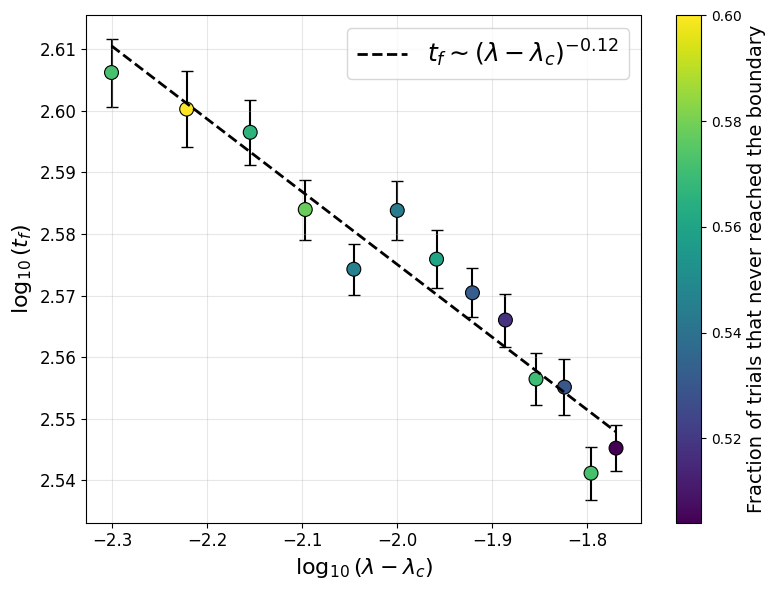}
    }
    \hfill
    \subfloat[
        Dependence of $t_f$ on system size $L$.
        \label{fig:tf_SIR_L}
    ]{
        \includegraphics[width=0.47\textwidth]
        {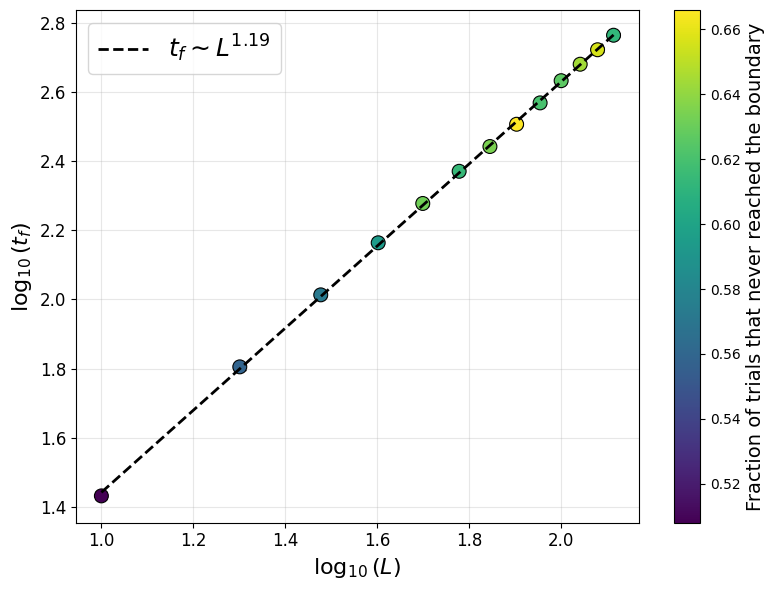}
    }
    \caption{
    Mean system traversal time $t_f$ for isolated SIR dynamics initialized with a single infected individual at the center of the lattice.
    (a) Double-logarithmic representation of $t_f$ as a function of the distance from the SIR epidemic threshold, $\lambda - \lambda_c$, for $\lambda > \lambda_c$, at fixed linear system size $L=100$.
    (b) Double-logarithmic representation of $t_f$ as a function of $L$ at the critical infection probability $\lambda = \lambda_c \approx 0.204$.
    For each parameter value, the mean traversal time is calculated only from the successful boundary-reaching realizations among $500$ independent stochastic realizations.
    Error bars indicate the corresponding statistical uncertainties, while the color scale shows the fraction of realizations in which the epidemic
    does not reach the system boundary within the observation interval of $5,000$ MCS.
    }
    \label{fig:tf_SIR}
\end{figure*}

\subsubsection{SIR dynamics}

We next perform the same analysis for the SIR model. 
In contrast to the SIS process, infected individuals that recover become permanently immune and therefore cannot participate in subsequent infection events. 
Consequently, the epidemic can propagate only once through a given portion of the susceptible population. 
Starting from a localized infectious seed, the dynamics therefore generates an outward-moving epidemic cluster that leaves recovered individuals behind.

Figure~\ref{fig:tf_SIR_delta} shows that the SIR traversal time follows the same qualitative trend as in the SIS model: 
Increasing the infection probability accelerates epidemic spreading and therefore reduces the time required for the epidemic cluster to reach the system boundary. 
Conversely, $t_f$ increases as $\lambda$ approaches the SIR epidemic threshold.
This dependence is, however, considerably weaker than for SIS dynamics. 
A linear fit in the double-logarithmic representation yields $t_f \sim (\lambda-\lambda_c)^{-0.118\pm0.009}$.
Thus, over the investigated supercritical parameter range, the characteristic propagation time of the SIR epidemic wave exhibits only a weak dependence on the distance from the threshold. 
As in the SIS case, this exponent characterizes front propagation and should not be directly identified with the temporal correlation exponent of the corresponding dynamic isotropic percolation (DIP) universality class.
The shorter traversal times observed for SIR dynamics can be understood from the absence of reinfection: 
Once an infected individual recovers, it becomes permanently immune. 
The infection front therefore advances primarily into previously susceptible regions instead of repeatedly spreading through already visited parts of the lattice. 
This produces a more clearly defined outward-moving epidemic wave. 
Consequently, epidemic activity reaches the system boundary considerably faster than in the corresponding SIS simulations.

The finite-size dependence of the SIR traversal time is shown in Fig.~\ref{fig:tf_SIR_L}. 
At the epidemic threshold, the data exhibit a remarkably clean algebraic dependence over the investigated range of system sizes, $t_f(L)\sim L^{1.190\pm0.005}$.
The measured slope corresponds to an effective dynamic exponent $z_{\mathrm{eff}} = 1.190\pm0.005$.
For two-dimensional dynamic isotropic percolation, the dynamic exponent equals the shortest-path fractal dimension of critical percolation clusters, for which high-precision numerical simulations yield $z = d_{\min} = 1.13077(2)$ \cite{ZhouEtAl2012}.
Our estimate is approximately $5\%$ larger than this asymptotic value.
Finite-size corrections and uncertainty in the location of the epidemic threshold can affect the traversal-time scaling over the accessible range of system sizes.
The substantially smaller exponent compared with the SIS result demonstrates that critical SIR epidemic clusters spread spatially more rapidly across the finite lattice.
The distinction between the SIS and SIR results reflects their very different underlying critical dynamics. 
The SIS transition belongs to the directed percolation universality class, whereas the SIR epidemic threshold is governed by dynamic isotropic percolation. 
In addition, their microscopic propagation mechanisms are qualitatively different: 
Repeated recovery and reinfection processes continually reorganize SIS activity, whereas the SIR epidemic develops as a single propagating wave that leaves an inactive immune population behind.

Taken together, the two sets of measurements demonstrate that the system traversal time provides a useful connection between critical spreading dynamics and the stabilization mechanism investigated in the coupled systems below. 
At the epidemic threshold, $t_f$ probes the anomalous relation between spatial and temporal scales through $t_f \sim L^z$. 
Above the threshold, its dependence on the control parameter instead reflects the emergence of a finite front velocity and its variation with the distance from criticality. 
Most importantly for the coupled systems studied in the following section, $t_f$ determines the characteristic time required for epidemic activity generated within an active region to propagate from its interior to the subsystem interface. 
The competition between this propagation time and the intrinsic extinction time of the neighboring vulnerable subsystem therefore provides a natural time-scale criterion for whether activity transfer can occur sufficiently rapidly to prevent stochastic extinction.
\section{Diffusively Coupled Spatially Distinct Subsystems}
\label{sec:coupled_systems}

Having characterized stochastic extinction and epidemic propagation in isolated systems, we now investigate how these stochastic processes compete when spatially distinct epidemic subsystems are diffusively coupled.
Specifically, we consider a vulnerable subsystem that, owing to its finite size and proximity to the epidemic threshold, is susceptible to stochastic extinction, but is connected to a neighboring region that generates more robust epidemic activity. 
Infection can propagate locally across the common interface between the two regions, allowing infectious clusters generated in the more robust subsystem to invade the extinction-prone region.

The central question is whether activity generated in the source subsystem can reach the neighboring vulnerable region and establish a sufficiently extended correlated infectious population to sustain a persistent epidemic. 
The initial activity in the vulnerable subsystem may already have disappeared before the invading front arrives. 
In this case, infections transmitted across the interface may reactivate the disease-free region and provide a new effective initial condition for the local epidemic dynamics.
Diffusive coupling of two subsystems therefore introduces an additional finite-size mechanism that may counteract stochastic extinction. 
Invading near-critical activity clusters generated in the more robust subsystem can effectively reset the vulnerable region with spatially correlated infectious configurations, thereby providing new favorable initial conditions for sustained epidemic spreading.

We first investigate this mechanism when both subsystems follow SIS dynamics, such that the more robust region provides a persistent source of epidemic activity. 
Since the critical spreading dynamics of the SIS model belongs to the directed percolation (DP) universality class, the infectious clusters generated in the active region exhibit the characteristic spatio-temporal properties associated with DP. 
We subsequently consider coupling to an SIR subsystem, for which the source activity is only transient and the critical spreading dynamics instead belongs to the dynamic isotropic percolation (DIP) universality class. 
Finally, we briefly address two diffusively coupled SIR regions and probe the conditions for a single epidemic wave in the more infectious subsystem to incite a sustained outbreak in the weaker patch.
Comparing these three cases therefore allows us not only to determine whether continued external forcing is necessary for stabilization, or whether a single sufficiently developed invading epidemic wave can induce long-lived activity in the vulnerable region, but also to investigate how infectious clusters originating from distinct universality classes influence the stabilization process and the spatially correlated population structures that emerge in the vulnerable subsystem.

\subsection{Stabilization of a vulnerable SIS subsystem by an active SIS region}
\label{sec:SIS_SIS}

\subsubsection{Density dynamics and vulnerable subsystem invasion}

We first consider two coupled SIS subsystems. 
The vulnerable region is assigned a fixed infection probability $\lambda_{\mathrm{v}} = 0.109$, which lies only slightly above the critical value $\lambda_c \approx 0.106$. 
Consequently, an isolated finite system set at these parameters remains highly susceptible to stochastic disease extinction. 
In the neighboring more infectious region, the infection probability $\lambda_{\mathrm{a}}$ is varied over the range $0.110 \leq \lambda_{\mathrm{a}} \leq 0.120$. 
These values place the source subsystem progressively farther into the active epidemic phase, thereby increasing its robustness against stochastic extinction and enhancing the propagation of epidemic activity toward the interface.

\begin{figure*}[t]
    \centering
    \subfloat[
        More robust SIS subsystem.
        \label{fig:active_SIS_density}
    ]{      \includegraphics[width=0.47\textwidth]
        {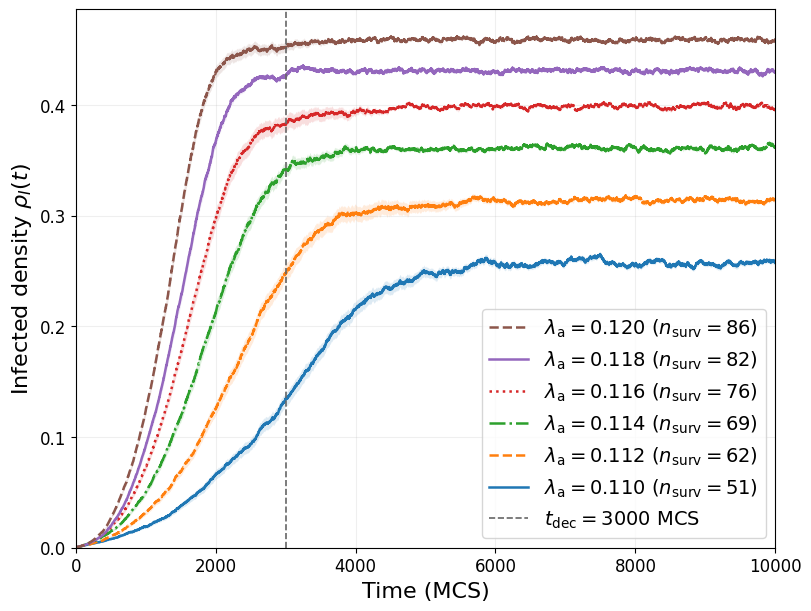}
    }
    \hfill
    \subfloat[
        Vulnerable SIS subsystem.
        \label{fig:coupled_SIS_density}
    ]{
        \includegraphics[width=0.47\textwidth]
        {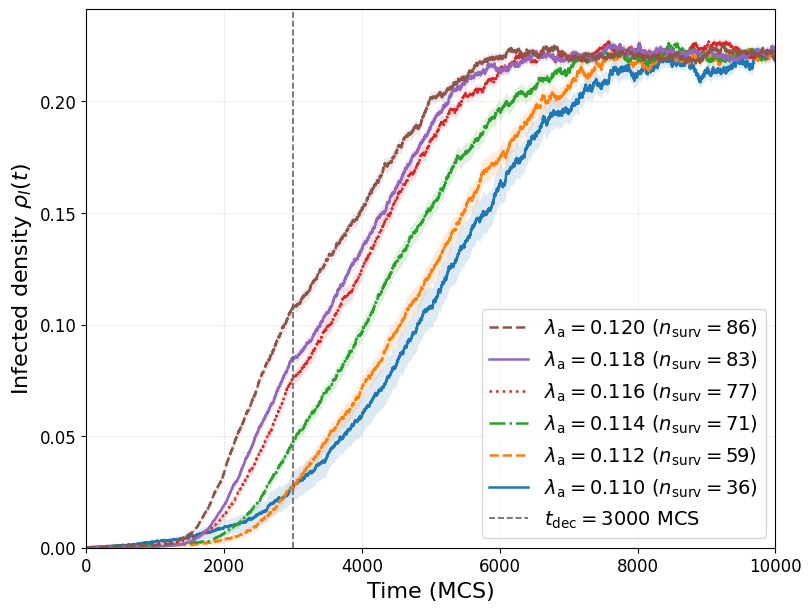}
    }
    \caption{
    Time evolution of the infected densities in two diffusively coupled SIS
    subsystems with common linear size $L=100$.
    The infection probability in the more robust subsystem $A$ is varied over the range $0.110\leq\lambda_{\mathrm{a}}\leq0.120$, while the infection probability in the vulnerable subsystem $B$ is fixed at $\lambda_{\mathrm{v}}=0.109$.
    The recovery probability is set to $\gamma=0.1$ in both regions. 
    Each subsystem is initialized with a single infected individual; the diffusive coupling is removed at $t_{\mathrm{dec}}=3,000$ MCS; and the results are obtained from $100$ independent stochastic realizations.
    For each infection probability, the densities are averaged over the realizations that remain active in the corresponding subsystem until $t=10,000$ MCS.
    Panel (a) shows the epidemic dynamics in the more robust SIS subsystem, while panel (b) displays the response of the vulnerable SIS subsystem.
    In panel (a), the quantity $n$ reported in the legend denotes the number of realizations in which subsystem $A$ remains active until the end of the observation interval at $t=10,000$ MCS.
    In panel (b), $n_{surv}$ denotes the number of realizations in which subsystem $B$ remains active until the same final time.
    For the investigated infection probabilities, every realization in which the more robust subsystem remains active also successfully restabilizes the vulnerable subsystem.
    }
    \label{fig:SIS_SIS_density}
\end{figure*}
Figure~\ref{fig:SIS_SIS_density} shows the epidemic dynamics in the two coupled SIS regions. 
The quantity $n$ reported in the legend denotes the number of stochastic realizations among the total $100$ simulation runs that remain active until the end of the observation window ($t=10,000$ MCS) for the corresponding subsystem.
In the more robust epidemic SIS region (Fig.~\ref{fig:active_SIS_density}), increasing $\lambda_{\mathrm{a}}$ enhances epidemic spreading and drives the system more rapidly toward its quasi-stationary active state. 
The characteristic rise of the infectious population therefore shifts systematically toward earlier times as $\lambda_{\mathrm{a}}$ is increased. 
In addition, the long-time quasi-stationary infected population grows with the infection probability, as expected for a system moved progressively farther into the active phase.
The survival statistics display the same trend: 
The number of realizations that remain active increases from $n=57$ for $\lambda_{\mathrm{a}}=0.110$ to $n=87$ for $\lambda_{\mathrm{a}}=0.120$, reflecting the increasing robustness of the source region against stochastic extinction.

The corresponding response of the vulnerable subsystem is shown in Fig.~\ref{fig:coupled_SIS_density}.
Infection clusters generated in the more robust region propagate toward the common interface and subsequently invade the vulnerable region \cite{SimulationMovies}.
As demonstrated by the system traversal time measurements in Sec.~\ref{sec:traversal_time}, increasing $\lambda_{\mathrm{a}}$ reduces the characteristic propagation time $t_f$.
Consequently, activity reaches the subsystem interface earlier for larger $\lambda_{\mathrm{a}}$, and epidemic growth in the vulnerable region begins at correspondingly earlier times. 
The vulnerable subsystem therefore also approaches its quasi-stationary active state more rapidly as the infection probability in the neighboring source region is increased.

A particularly notable feature is the comparison of the survival counts in the two subsystems. For the weakest source considered, $\lambda_{\mathrm{a}}=0.110$, $57$ realizations remain active in the more robust subsystem, whereas only $29$ corresponding realizations remain active in the vulnerable region. 
Thus, at this infection probability, survival of the source region does not necessarily imply successful stabilization of the neighboring subsystem. 
For all larger infection probabilities $\lambda_{\mathrm{a}}\geq0.112$ shown, however, the number of surviving realizations is identical in both regions. 
Hence, whenever the more robust SIS subsystem survives and generates sufficiently strong invading activity, the vulnerable subsystem is also successfully restabilized. 
This observation provides direct evidence that sufficiently rapid propagation of activity from the source region can suppress stochastic extinction in the neighboring vulnerable subsystem.

Importantly, the results discussed below demonstrate that this stabilization does not require a continuous influx of infected individuals indefinitely.
For the simulations shown here, the two subsystems remain coupled until $t_{\mathrm{dec}} = 3,000$ MCS, after which the coupling is removed.
Once the invading activity has penetrated sufficiently far into the vulnerable subsystem and established a spatially extended correlated active population, the vulnerable region may subsequently remain active autonomously even after the two subsystems are decoupled.

\subsubsection{Spatial correlation dynamics across the system interfaces}
\begin{figure*}[t]
    \centering
    \subfloat[
        Correlation length parallel to the interface.
        \label{fig:xi_parallel_interface}
    ]{
        \includegraphics[width=0.47\textwidth]
        {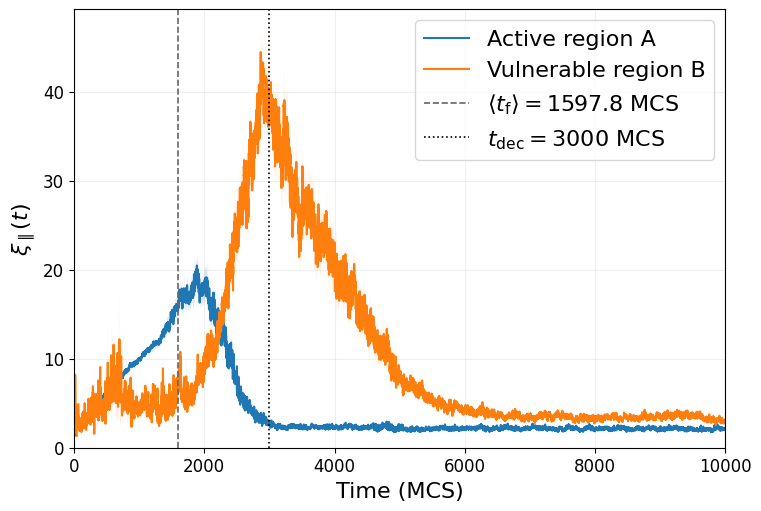}
    }
    \hfill
    \subfloat[
        Correlation length perpendicular to the interface.
        \label{fig:xi_perpendicular_interface}
    ]{
        \includegraphics[width=0.47\textwidth]
        {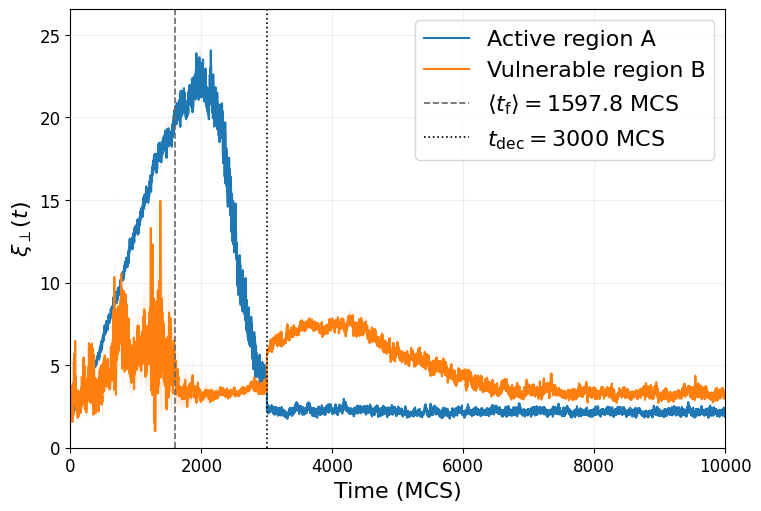}
    }
    \caption{
    Time evolution of the correlation lengths parallel and perpendicular to the interface in the more robust SIS subsystem $A$ and the vulnerable SIS subsystem $B$.
    Both subsystems have linear size $L=100$, with infection probabilities $\lambda_{\mathrm{a}}=0.115$ and $\lambda_{\mathrm{v}}=0.109$ and recovery probability $\gamma=0.1$.
    Each subsystem is initialized with a single infected individual; their
    coupling is removed at $t_{\mathrm{dec}}=3,000$ MCS.
    Panel (a) shows the correlation length
    $\xi_{\parallel}(t)$ parallel to the common interface; panel (b) displays the correlation length $\xi_{\perp}(t)$ perpendicular to the interface.
    The dashed line marks the mean first arrival at an interface of subsystem $A$ ($t  = 1,597.8$ MCS), and the dotted line marks decoupling at $t_{\mathrm{dec}}=3,000$ MCS.
    The correlation lengths are extracted from exponential fits to the equal-time spatial correlation functions. Following this arrival, extended spatial correlations develop in the vulnerable subsystem, with a stronger enhancement parallel than perpendicular to the interface.
    }
    \label{fig:xi_interface}
\end{figure*}
To gain further insight into the microscopic mechanism responsible for the stabilization of the vulnerable subsystem, we next investigate the evolution of spatial correlations during the coupled dynamics. 
While the density measurements presented in the previous section demonstrate that diffusive coupling prevents stochastic extinction through persistent spreading of active clusters into the vulnerable region, they do not reveal spatial information how epidemic activity is transmitted across the interface. 
For this purpose, we compute the equal-time spatial correlation function separately within the active and vulnerable subsystems and analyze its evolution both parallel (longitudinal) and perpendicular to the coupling interface.

We consider separations $r$ parallel ($\alpha=\parallel$) and perpendicular ($\alpha=\perp$) to the interfaces. 
For each realization and subsystem $X\in\{A,B\}$, we compute
\begin{equation}
C_\alpha^{X}(r,t) = \big\langle \eta_i(t) \,\eta_j(t) \big\rangle_{\mathrm{pairs}} - \big\langle n_I^{X}(t)\big\rangle^2 \, ,
\end{equation}
where $\eta_i(t)$ is one if site $i$ is infected and zero otherwise, $n_I^{X}(t)$ is the instantaneous infected density in subsystem $X$, and the average is taken over the valid pairs of sites in $X$ separated by $r$ in direction $\alpha$.
The pair average parallel to the interfaces uses periodic boundary conditions at all times. 
While the subsystems are coupled, the perpendicular average includes only pairs whose separation can be realized within the same subsystem without wrapping across its edges.
At and after decoupling, perpendicular separations are evaluated periodically within each subsystem.

The corresponding correlation lengths $\xi_{\parallel}(t)$ and $\xi_{\perp}(t)$ are extracted from the spatial decay of the correlation functions measured along (parallel to) and perpendicular to the subsystem interface.
At each time $t$, we first average the correlation function over $100$ realizations in which activity from the active subsystem successfully reaches the interface. 
We then assume an approximately exponential decay over the selected fitting range at sufficiently large distances $r$,
\begin{equation}
C_{\alpha}(r,t) \simeq A_{\alpha}(t) \, e^{- r / \xi_{\alpha}(t)} \, , 
\quad \alpha \in \{\parallel,\perp\} \, .
\end{equation}
Taking the logarithm yields $\ln C_{\alpha}(r,t) \simeq \ln A_{\alpha}(t) - r / \xi_{\alpha}(t)$.
Thus, a linear fit of $\ln C_{\alpha}(r,t)$ as a function of distance $r$ produces a slope $- 1 / \xi_{\alpha}(t)$, from which the correlation lengths are straightforwardly obtained and the results displayed in Fig.~\ref{fig:xi_interface}.
Here, the vertical dashed line indicates the mean boundary arrival time $t_{\mathrm{arr}} = 1769.1 \pm 40.4 \ {\rm MCS}$ which marks the average time at which epidemic activity first reaches the interfaces between both subsystems.

Figure~\ref{fig:xi_interface} shows the temporal evolution of the correlation lengths parallel and perpendicular to the interface in the active and vulnerable SIS subsystems.
Before the arrival of the epidemic activity at the interface, the active subsystem develops increasingly long-ranged spatial correlations as the infection spreads away from its initial seed.
The correlation lengths grow continuously and reach their maxima close to the mean arrival time at the interface.
In contrast, the vulnerable subsystem remains only weakly correlated, with shorter correlation lengths, prior to the arrival of epidemic activity. 
Immediately after the interface is reached, however, the longitudinal correlation length within the vulnerable subsystem increases rapidly and substantially exceeds its pre-arrival range. 
This behavior indicates that the activity entering the vulnerable subsystem is not composed of isolated infected individuals, but rather consists of spatially extended correlated epidemic clusters that preserve their spatial coherence while propagating across the interface.

\begin{figure*}[t]
    \centering
    \subfloat[
        Active subsystem: correlations parallel to the interfaces.
        \label{fig:C_parallel_active}
    ]{
        \includegraphics[width=0.47\textwidth]
        {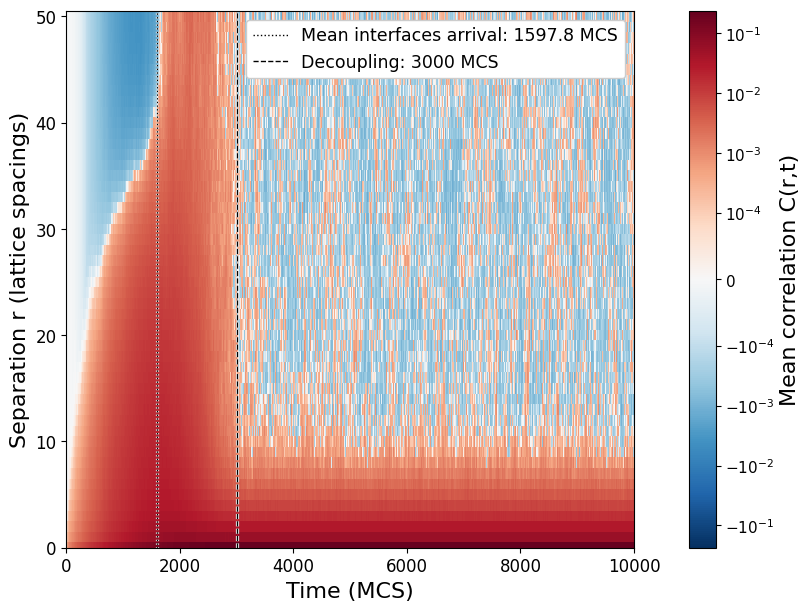}
    }
    \hfill
    \subfloat[
        Active subsystem: correlations perpendicular to the interfaces.
        \label{fig:C_perpendicular_active}
    ]{
        \includegraphics[width=0.47\textwidth]
        {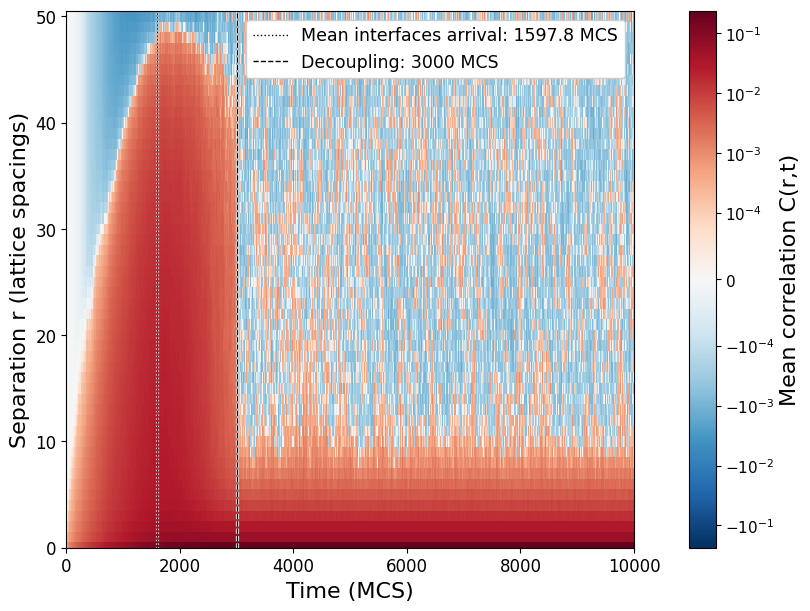}
    }
    \vspace{0.4cm}
    \subfloat[
        Vulnerable subsystem: correlations parallel to the interfaces.
        \label{fig:C_parallel_vulnerable}
    ]{
        \includegraphics[width=0.47\textwidth]
        {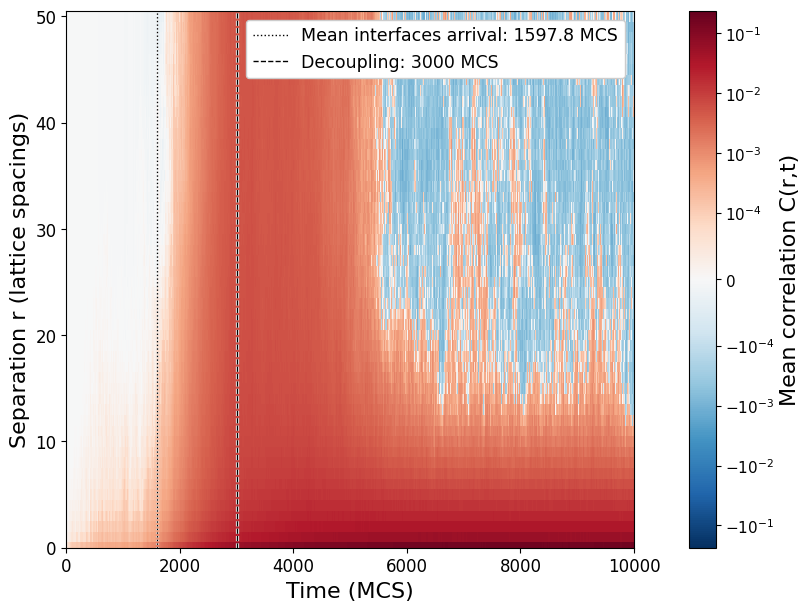}
    }
    \hfill
    \subfloat[
        Vulnerable subsystem: correlations perpendicular to the interfaces.
        \label{fig:C_perpendicular_vulnerable}
    ]{
        \includegraphics[width=0.47\textwidth]
        {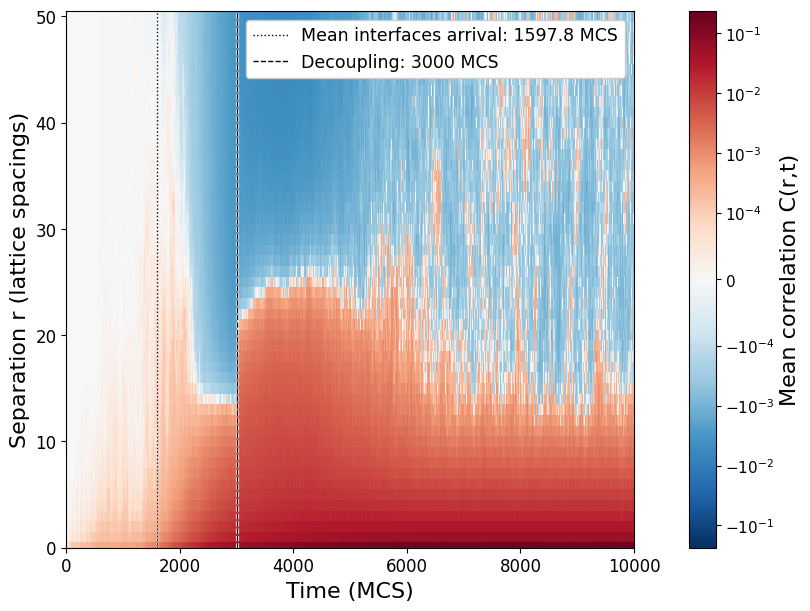}
    }
    \caption{
    Space-time evolution of the equal-time spatial correlation functions $C_\alpha(r,t)$ in two diffusively coupled SIS subsystems.
    The left column shows correlations measured parallel to the common interfaces, while the right column shows correlations measured perpendicular to the interfaces.
    The upper row (a,b) presents the correlations in the more robust subsystem $A$; the lower row (c,d) displays the correlations in the vulnerable subsystem $B$.
    Both subsystems have linear size $L=100$, with infection probabilities $\lambda_{\mathrm{a}}=0.115$ and $\lambda_{\mathrm{v}}=0.109$ and recovery probability $\gamma=0.1$. 
    Each subsystem is initialized with a single infected individual; their diffusive coupling is removed at $t_{\mathrm{dec}}=3,000$ MCS. 
    The correlations are averaged over $100$ independent realizations. 
    The dotted line marks the mean first arrival at an interface of subsystem $A$ ($1,597.8$ MCS), and the dashed line marks decoupling at $t_{\mathrm{dec}}=3,000$ MCS. 
    All four panels utilize the same symmetric logarithmic color scale, with a linear interval close to zero; red and blue denote positive and negative correlations, respectively. 
    Following the arrival of epidemic activity at the interface, extended positive correlations develop in the vulnerable region, particularly parallel to the interface.
    }
    \label{fig:heatmap_corr}
\end{figure*}
Interestingly, the enhancement is considerably stronger in the longitudinal direction than in the perpendicular direction. 
Since the coupling occurs through a planar interface, epidemic activity propagates preferentially normal to the interface, producing elongated correlated structures that extend deep into the vulnerable region. 
The ensuing pronounced anisotropy provides direct evidence that the stabilization mechanism is associated with the directional transmission of spatial correlations generated within the active subsystem. 

\begin{figure*}[t]
    \centering
    \subfloat[
        More robust SIS subsystem.
        \label{fig:active_SIS_density_size}
    ]{
        \includegraphics[width=0.47\textwidth]
        {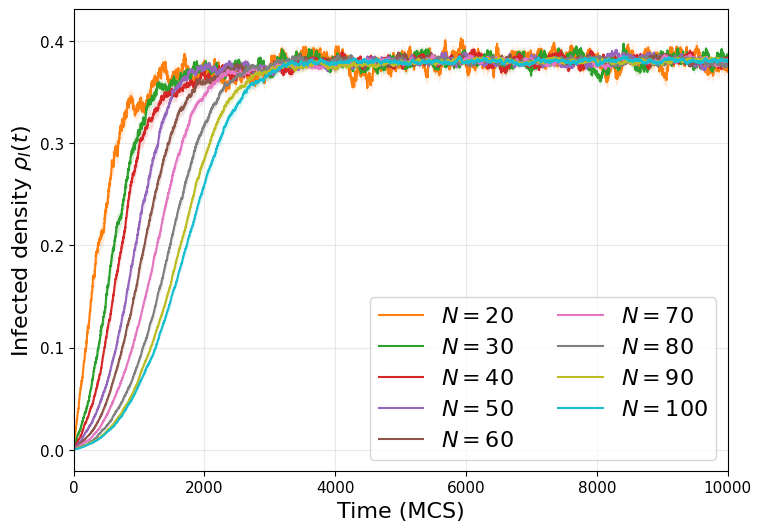}
    }
    \hfill
    \subfloat[
        Vulnerable SIS subsystem.
       \label{fig:vulner_SIS_density_size}
    ]{
        \includegraphics[width=0.47\textwidth]
        {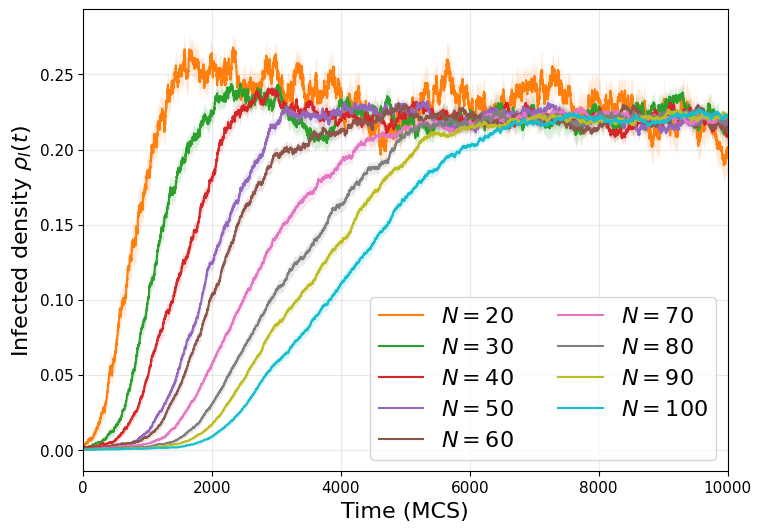}
    }
    \caption{
    Time evolution of the infected densities in two diffusively coupled SIS subsystems for different common linear system sizes $L$. The infection probabilities are fixed at $\lambda_{\mathrm{a}}=0.115$ in the more robust subsystem and $\lambda_{\mathrm{v}}=0.109$ in the vulnerable subsystem, with the recovery probability set to $\gamma=0.1$ in both regions.
    Each subsystem is initialized with a single infected individual; the coupling is removed at $t_{\mathrm{dec}}=3,000$ MCS; and the results are obtained from $100$ independent stochastic realizations for each system size.
    Panel (a) shows the infected density in the more robust subsystem, averaged over realizations in which that subsystem remains active until $t=10,000$ MCS.
    Panel (b) shows the infected density in the vulnerable subsystem, averaged over realizations in which that subsystem remains active during that same time interval.
    Increasing $L$ delays both the propagation of epidemic activity to the interface and the subsequent reactivation of the vulnerable subsystem.
    Once activity becomes established, the surviving realizations approach approximately the same quasi-stationary infected density.
    }
    \label{fig:SIS_SIS_density_size}
\end{figure*}
At decoupling ($t_{\mathrm{dec}}=3,000$ MCS), the perpendicular correlation length $\xi_{\perp}(t)$ shows a step-like decrease in the active subsystem and a similar sharp increase in the vulnerable subsystem, Fig.~\ref{fig:xi_interface}(b). 
Prior to decoupling, the perpendicular correlation function is evaluated using non-wrapping site pairs within each subsystem. 
When periodic boundary conditions are imposed in this direction, sites near opposite edges of the same subsystem can contribute to the correlation
function at short separations. 
The resulting shift in the fitted correlation length depends on the spatial distribution of infected sites in each subsystem.

To visualize this process directly, Fig.~\ref{fig:heatmap_corr} presents the time evolution of the equal-time spatial correlation functions measured in both subsystems.
The displayed correlation function heat maps provide a complementary picture of the stabilization process.
Prior to the arrival time, significant correlations are confined almost entirely to the active subsystem. 
Following the arrival of epidemic activity at the interface, long-range correlations emerge inside the vulnerable subsystem, particularly in the longitudinal direction, parallel to the interfaces. 
The correlation amplitude simultaneously extends to larger spatial separations, demonstrating that the invading activity carries correlated epidemic structures into the previously extinction-prone region.
The change in the perpendicular correlation pattern at $t_{\mathrm{dec}}$ coincides with the introduction of periodic pair sampling in that direction. 
The altered set of sampled pairs can therefore contribute to the apparent discontinuity at decoupling. 
The parallel pair convention is the same throughout, and the extended positive correlations along the interfaces persist
for a finite time after the coupling is removed.

As time proceeds, the initially extended invasion cluster breaks into multiple smaller activity centers, typically only a few lattice sites across, in both subsystems, consistent with the decrease of the correlation lengths displayed in Fig.~\ref{fig:xi_interface}. 
The system subsequently enters a quasi-stationary regime characterized by short-ranged, spatially separated infection clusters, which remain slightly larger in the vulnerable region than in the more robust subsystem. 
The simulation movie available at Ref.~\cite{SimulationMovies} shows the fragmentation and subsequent reorganization of the invading activity.

These observations strongly support the interpretation that stabilization is not simply caused by a continuous influx of infected individuals across the interface. 
Instead, the active subsystem transfers large near-critical correlated epidemic clusters into the vulnerable region. 
Once these spatially and temporally correlated structures become sufficiently extended, they establish favorable initial conditions for sustained epidemic spreading and thereby strongly suppress stochastic extinction even after the incoming activity has crossed the interface.

\subsubsection{Stabilization mechanism: finite-size dependence}
\label{sec:SIS_SIS_finite_size}

We next investigate how the stabilization mechanism depends on the linear size $L$ of the coupled subsystems. 
For this purpose, we fix the infection probabilities at $\lambda_{\mathrm{a}}=0.115$ in the more robust SIS subsystem and $\lambda_{\mathrm{v}}=0.109$ in the vulnerable SIS subsystem, while varying the common linear system size.
The recovery probability remains fixed at $\gamma=0.1$. 
As before, each subsystem is initialized with a single infected individual, the two regions remain coupled until $t_{\mathrm{dec}}=3,000$ MCS, and the dynamics are followed until $t=10,000$ MCS.

Figure~\ref{fig:SIS_SIS_density_size} shows the time evolution of the infected density in the two subsystems for different system sizes. 
In the more robust region, {\em c.f.} Fig.~\ref{fig:active_SIS_density_size}, the time required to reach the quasi-stationary state increases systematically with $L$. 
This behavior is expected because the epidemic starts from a localized seed and must spread over a progressively larger spatial region before a system-spanning active population is established. 
The associated time scale is therefore controlled by epidemic propagation across the finite lattice.
Despite this size-dependent transient, all curves approach approximately the same asymptotic infected density, $\rho_I^{\mathrm{qs}}\approx 0.38$. 
Thus, over the range of system sizes considered here, the quasi-stationary density of the active region is essentially independent of $L$ and instead is primarily determined by its fixed infectivity $\lambda_{\mathrm{a}}=0.115$. 
The larger fluctuations visible for the smaller systems result from the lower number of lattice sites and the correspondingly stronger finite-size fluctuations.

The vulnerable region displays a similar shift of the growth dynamics toward later times as $L$ increases, see Fig.~\ref{fig:vulner_SIS_density_size}. 
For small systems, activity originating in the more robust region has a shorter distance to traverse before reaching the interface and spreading throughout the vulnerable region.
Consequently, the increase of the infected density in the vulnerable subsystem begins earlier for smaller $L$. 
In larger systems, the longer propagation distance delays the arrival of invading activity and the subsequent growth of the infected population.

It is important, however, that the density curves in Fig.~\ref{fig:SIS_SIS_density_size} are conditioned on the realizations that remain active until the end of the observation window. 
Hence, they describe the temporal evolution following a {\em successful} invasion, but do not measure how frequently such an invasion effectively stabilizes the vulnerable subsystem. 
Although the vulnerable region is reached earlier in a small system, successful long-time stabilization is considerably less likely there, as demonstrated below.

\begin{figure}[t]
    \centering
    \includegraphics[width=0.95\columnwidth]
    {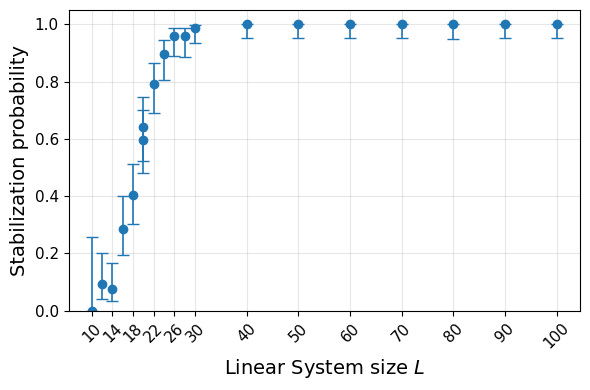}
    \caption{
    Stabilization probability $P_{\mathrm{stab}}(L)$ as a function of the common linear system size $L$.
    The stabilization probability is defined as the conditional probability that the vulnerable SIS subsystem $B$ remains active until $t=10,000$ MCS, given that the more robust SIS subsystem $A$ remains active until the same final time.
    The infection and recovery probabilities are $\lambda_{\mathrm{a}}=0.115$, $\lambda_{\mathrm{v}}=0.109$, and $\gamma=0.1$.
    Both subsystems are initialized with a single infected individual; the coupling is removed at $t_{\mathrm{dec}}=3,000$ MCS.
    Each data point is obtained from $100$ independent stochastic realizations.
    The conditional stabilization probability increases rapidly with $L$ and reaches $1$ for $L\geq50$ within the sampled ensemble.
    }
    \label{fig:SIS_SIS_stabilization_size}
\end{figure}
To quantify the reliability of the stabilization process, we measure the conditional survival probability
\begin{equation}
P_{\mathrm{stab}}(L) = P\!\left( B\ {\rm survives} \,\middle|\, A\ {\rm survives} \right),
\label{eq:stabilization_probability_size}
\end{equation}
where $A$ indicates the more robust SIS region and $B$ the vulnerable SIS region. 
This quantity isolates the effectiveness of activity transfer from the probability that the source region itself survives. 
In the simulations, it is estimated from the fraction of realizations with a surviving region $A$ in which region $B$ also remains active until the end of the observation window.
As shown in Fig.~\ref{fig:SIS_SIS_stabilization_size}, $P_{\mathrm{stab}}(L)$ is close to zero for the smallest systems and then increases rapidly with system size. 
For $L\geq 50$, the measured conditional probability is unity: 
Within the available ensemble of $100$ realizations, whenever the more robust subsystem survives, the vulnerable subsystem also persists. 
The transition between these regimes is continuous rather than abrupt, with the strongest size dependence occurring approximately over $14\lesssim L\lesssim 26$.

This result reveals an important distinction between the time at which activity first reaches the vulnerable region and the probability that this infectious activity produces lasting epidemic stabilization. 
Smaller systems are traversed more quickly, but they also experience stronger demographic fluctuations and permit only spatially restricted infection clusters. 
Activity transferred across the interface may therefore disappear before it develops into a sufficiently robust correlated infectious population. 
Increasing the system size provides more  space for the invading clusters to grow and reduces their susceptibility to a system-wide fluctuation into the absorbing state, thus strongly increasing the probability of successful stabilization.

\begin{figure}[t]
    \centering
    \includegraphics[width=0.95\columnwidth]
    {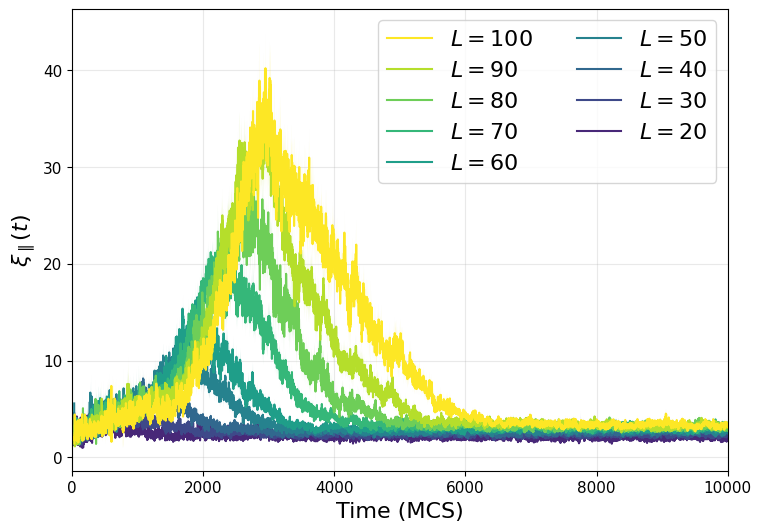}
    \caption{
    Time evolution of the correlation length $\xi_{\parallel}(t)$ parallel to the interface in the vulnerable SIS subsystem for different common linear system sizes $L$.
    The infection probabilities are fixed at $\lambda_{\mathrm{a}}=0.115$ in the more robust subsystem and at $\lambda_{\mathrm{v}}=0.109$ in the vulnerable subsystem, with the recovery   probability set to $\gamma=0.1$ in both regions.
    Each subsystem is initialized with a single infected individual; the diffusive coupling is removed at $t_{\mathrm{dec}}=3,000$ MCS; and the results are obtained from $100$ independent stochastic realizations for each system size.
    The correlation lengths are extracted from exponential fits to the equal-time spatial correlation functions.
    Increasing $L$ produces a larger and later transient maximum in $\xi_{\parallel}(t)$, consistent with the formation of increasingly extended correlated populations and the enhanced stabilization probability observed in larger systems.
    The correlation functions are averaged over realizations in which the vulnerable subsystem remains active until the end of the observation interval.
    }
    \label{fig:SIS_SIS_xi_parallel_size}
\end{figure}

The corresponding finite-size evolution of the correlation length parallel to the interface is shown in Fig.~\ref{fig:SIS_SIS_xi_parallel_size}. 
For the smallest systems, $\xi_{\parallel}(t)$ remains limited to only a few lattice spacings.
Consequently, the activity entering the vulnerable subsystem cannot develop a spatially extended correlated structure before finite-size fluctuations can eliminate it. 
This behavior is consistent with the low stabilization probability observed at small $L$.
As the system size increases, the longitudinal correlation length develops a progressively larger transient maximum. 
The peak also shifts toward later times because epidemic activity requires longer to reach the interface and spread through the larger vulnerable subsystem. 
For the largest systems, the maximum correlation length extends over a substantial fraction of the linear system size. 
The invading activity can consequently generate an extended correlated infected population rather than merely a small, localized collection of infected sites.
Such a near-critical infectious population cluster constitutes a more favorable effective initial condition for autonomous epidemic persistence after the subsystems are decoupled.

At sufficiently late times, $\xi_{\parallel}(t)$ decreases from its transient maximum and approaches a much smaller quasi-stationary value. 
This decline does not indicate a loss of epidemic activity. 
Rather, it reflects the fragmentation of the initially extended invasion structure into the smaller infection clusters characteristic of the quasi-stationary SIS state. 
The finite-size dependence of the transient correlation length therefore supports the interpretation inferred from Fig.~\ref{fig:SIS_SIS_stabilization_size}: 
Successful persistent stabilization requires the invading activity to establish a sufficiently extended correlated population in the vulnerable subsystem. 
Small systems strongly restrict the development of such structures, whereas larger systems allow them to form and thereby render reactivation and stabilization increasingly reliable.

\subsubsection{Effect of the coupling duration on activity re-emergence and stabilization}

The previous correlation analysis suggests that stabilization of the vulnerable subsystem is associated with the transmission and subsequent development of spatially extended correlated activity across the interface. 
To further investigate this connection, we systematically vary the duration for which the two subsystems remain diffusively coupled. 
Specifically, both regions are separated and hence the coupling switched off at a prescribed decoupling time $t_{\mathrm{dec}}$, which is varied from $2,000$ to $3,000$ MCS in increments of $100$ MCS. 
For each value of $t_{\mathrm{dec}}$, we perform $100$ independent stochastic realizations.
To quantify the ability of the active subsystem $A$ to stabilize the vulnerable subsystem $B$, we define the stabilization probability as the conditional survival probability
\begin{equation}
P_{\mathrm{stab}}(t_{\mathrm{dec}}) = 
P\!\left( B\ {\rm survives} \,\middle|\, A\ {\rm survives} \right) \, \approx \,
\frac{ N_{A\cap B}^{\mathrm{surv}}}{N_A^{\mathrm{surv}}} \ ,
\label{eq:stabilization_probability}
\end{equation}
where $N_A^{\mathrm{surv}}$ denotes the number of realizations in which the active subsystem survives and $N_{A\cap B}^{\mathrm{surv}}$ denotes the number of realizations in which activity persists in both subsystems. 
This quantity therefore directly measures the probability that a surviving active region successfully stabilizes the vulnerable region.

\begin{figure}[t]
    \centering
    \includegraphics[width=0.95\columnwidth]
    {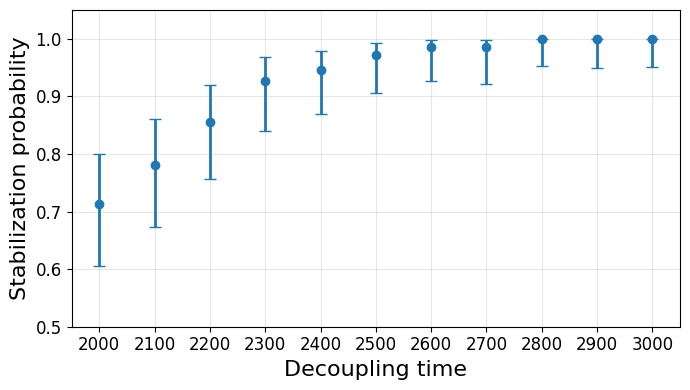}
    \caption{
    Stabilization probability $P_{\mathrm{stab}}$ as a function of the   decoupling time $t_{\mathrm{dec}}$.
    The stabilization probability is defined as the conditional probability that the vulnerable SIS subsystem $B$ remains active until $t=10,000$ MCS, provided that the more robust SIS subsystem $A$ remains active until the same final time.
    Both subsystems have linear size $L=100$, with infection and recovery probabilities $\lambda_{\mathrm{a}}=0.115$, $\lambda_{\mathrm{v}}=0.109$, and $\gamma=0.1$.
    Each subsystem is initialized with a single infected individual.
    All data points are obtained from $100$ independent stochastic realizations;
    displayed error bars indicate the statistical uncertainty of the estimated conditional probability.
    Increasing the coupling duration raises its average from $P_{\mathrm{stab}} \approx 0.7$ at $t_{\mathrm{dec}}=2,000$ MCS to $P_{\mathrm{stab}} \approx 1$ for $t_{\mathrm{dec}}\gtrsim2,700$ MCS.
    }
    \label{fig:stabilization_decoupling}
\end{figure}
Figure~\ref{fig:stabilization_decoupling} shows how the stabilization probability increases systematically with the duration of the coupling. 
For the earliest decoupling time $t_{\mathrm{dec}}=2,000$ MCS, only about $70\%$ of the realizations in which subsystem $A$ remains active also result in persistent disease spreading in region $B$. 
Increasing the coupling duration substantially enhances the probability of stabilization. 
For $t_{\mathrm{dec}}\gtrsim2,700$ MCS, the conditional survival probability approaches unity, and for the largest decoupling times essentially every realization with a surviving active patch also yields a persistent epidemic in the vulnerable subsystem.

This characteristic crossover time is of the same order as, but larger than the measured mean system traversal time $t_f=t_{\mathrm{arr}}=1,769.1\pm40.4$ MCS shown in Fig.~\ref{fig:xi_interface}. 
The distinction between these two time scales is physically important: 
The traversal time marks the first arrival of epidemic activity at the subsystem interface, whereas the decoupling time required for nearly certain stabilization is approximately $900$ MCS later. 
Thus, the first arrival of infected individuals is not sufficient to
guarantee survival of the vulnerable subsystem. Following boundary arrival, the two regions must remain coupled long enough for the incoming activity to penetrate the vulnerable region and develop into a spatially extended correlated infectious nucleus.
Once this seed of epidemic activity has become sufficiently established, subsystem $B$ can sustain the infection autonomously even after the coupling is removed.

\begin{figure}[t]
    \centering
    \includegraphics[width=0.95\columnwidth]
    {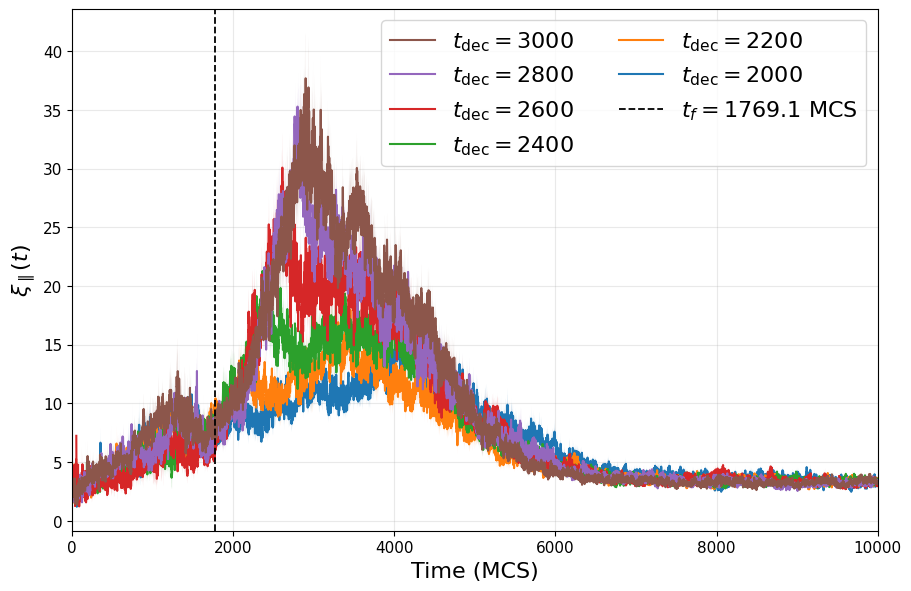}
    \caption{
    Time evolution of the correlation length   $\xi_{\parallel}(t)$ parallel to the interface in the vulnerable SIS subsystem for different decoupling times $t_{\mathrm{dec}}$.
    Both subsystems have linear size $L=100$, with infection and recovery probabilities $\lambda_{\mathrm{a}}=0.115$, $\lambda_{\mathrm{v}}=0.109$, and $\gamma=0.1$; they are each initialized with a single infected individual. 
    The results are obtained from $100$ independent stochastic realizations for each value of $t_{\mathrm{dec}}$.
    The correlation lengths are extracted from exponential fits to the equal-time spatial correlation functions.
    The mean system traversal time $t_f=1769.1\pm40.4$ MCS marks the average first arrival of epidemic activity at the interface.
    Since all displayed decoupling times occur after this mean arrival time, increasing $t_{\mathrm{dec}}$ provides a longer post-arrival interval during which the invading activity can spread through the vulnerable subsystem and develop an increasingly large maximum for $\xi_{\parallel}(t)$.
    The averages are conditioned on realizations in which the vulnerable region remains active until $t=10,000$ MCS.
    The vertical dashed line marks the mean system traversal time $t_f=1,769.1\pm40.4$ MCS, corresponding to the average first arrival of epidemic activity at the interface.
    }
    \label{fig:xi_parallel_decoupling}
\end{figure}

To connect this survival behavior directly with the development of spatial correlations, we next examine the time dependence of the longitudinal correlation length $\xi_{\parallel}(t)$ in the vulnerable subsystem for different values of $t_{\mathrm{dec}}$.
Figure~\ref{fig:xi_parallel_decoupling} reveals a clear connection between the coupling duration and the spatial extent of the correlated activity established inside the vulnerable subsystem. 
The mean system traversal time $t_f=1,769.1\pm40.4$ MCS marks the average first arrival of epidemic activity at the interface. 
Consequently, the investigated decoupling times provide post-arrival growth intervals ranging from approximately $230$ MCS for $t_{\mathrm{dec}}=2,000$ MCS to approximately $1,230$ MCS for $t_{\mathrm{dec}}=3,000$ MCS.

For early decoupling, the incoming activity has only a short interval in which to spread inside the vulnerable subsystem, and the parallel correlation length increases only moderately. As $t_{\mathrm{dec}}$ is increased, the maximum of $\xi_{\parallel}(t)$ grows substantially because the correlated activity has more time to propagate along the interface and become established within the vulnerable region. 
The development of these extended correlations during the post-arrival interval explains why the stabilization probability does not become close to unity immediately at $t_f$, but only at the later decoupling time $t_{\mathrm{dec}}\approx2,700$ MCS.

The simultaneous increase of $\xi_{\parallel}$ and $P_{\mathrm{stab}}$ provides a direct connection between the spatial correlation structure generated by the coupling and the subsequent survival of the vulnerable subsystem. 
In particular, the approach of $P_{\mathrm{stab}}$ toward unity for sufficiently large $t_{\mathrm{dec}}$ coincides with the development of strongly extended longitudinal correlations in subsystem $B$. 
These results therefore support the interpretation that successful stabilization requires not merely the arrival of infected individuals from the active subsystem, but sufficient time for the incoming activity to establish an extended \textit{correlated} population nucleus within the vulnerable region. 
Once such correlated activity centers have developed over a sufficiently large spatial scale, the vulnerable subsystem becomes considerably more resistant to stochastic extinction and can remain active even after the two regions are decoupled.
\subsection{Stabilization of a vulnerable SIS subsystem by a transient SIR epidemic}
\label{sec:SIR_SIS}

\subsubsection{Density dynamics and transient invasion}

\begin{figure*}[t]
    \centering
    \subfloat[
        Transient epidemic dynamics in the active SIR subsystem.
        \label{fig:active_SIR_density}
    ]{
        \includegraphics[width=0.47\textwidth]
        {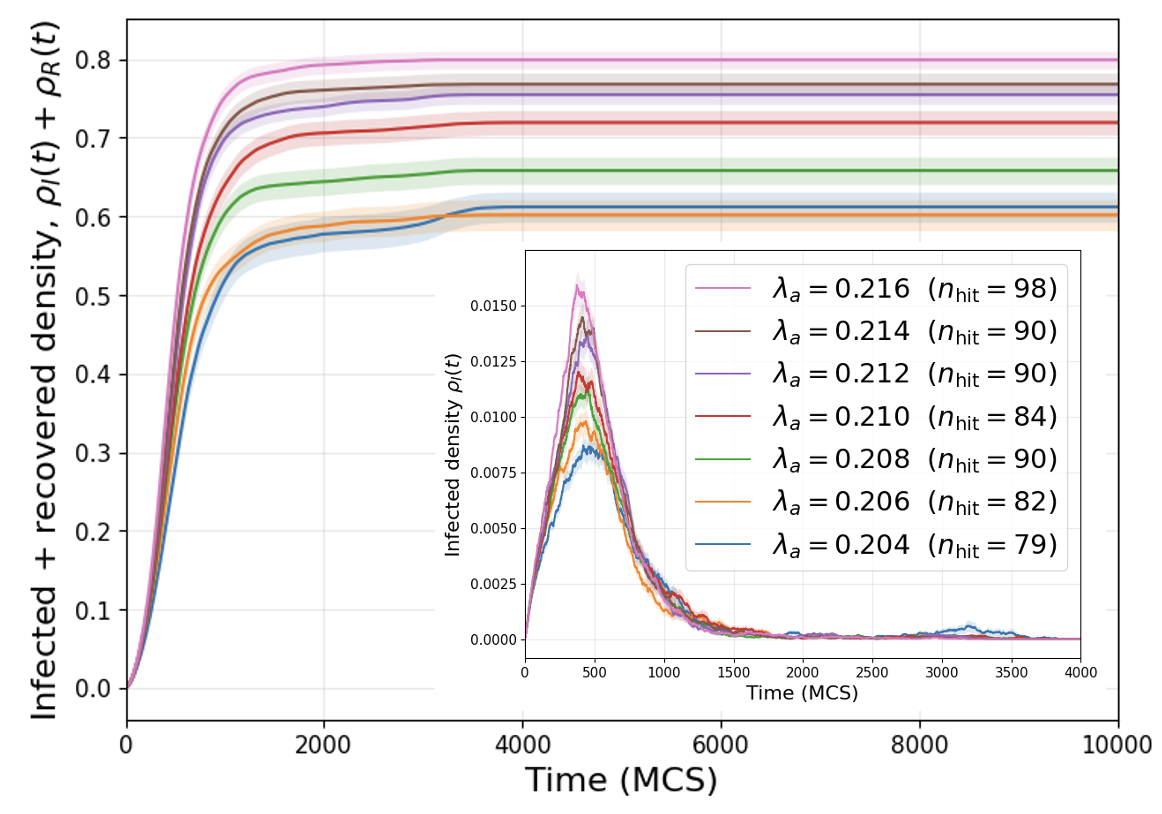}
    }
    \hfill
    \subfloat[
        Response of the vulnerable SIS subsystem.
        \label{fig:coupled_SIR_density}
    ]{
        \includegraphics[width=0.47\textwidth]
        {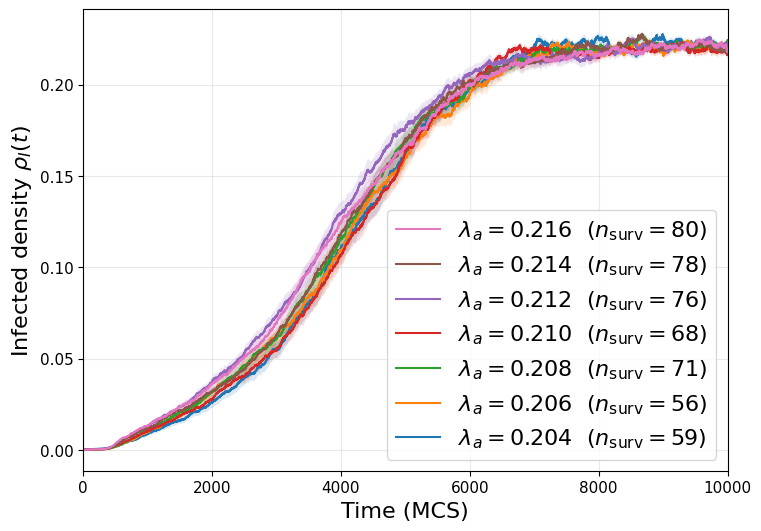}
    }
    \caption{
    Epidemic dynamics in a coupled SIR--SIS system composed of two subsystems with linear size $L=100$.
    The infection probability in the active SIR region is varied over the range $0.204\leq\lambda_{\mathrm{a}}\leq0.216$, above the SIR epidemic threshold $\lambda_c^{\mathrm{SIR}}\approx0.203$, while the infectivity in the vulnerable SIS region is fixed at the sub-threshold value $\lambda_{\mathrm{v}}=0.109$.
    The recovery probability is $\gamma=0.1$ in both subsystems, which are also both initialized with a single infected individual.
    The results are obtained from $100$ independent stochastic realizations for each value of $\lambda_{\mathrm{a}}$.
    Panel (a) shows the total affected density $\rho_I^{(A)}(t)+\rho_R^{(A)}(t)$ in the active SIR subsystem; the inset displays the corresponding transient infectious density $\rho_I^{(A)}(t)$.
    Panel (b) displays the infected density $\rho_I^{(B)}(t)$ in the vulnerable SIS subsystem.
    The quantity $n$ reported in the legend of panel (a) indicates the number of realizations in which the transient SIR epidemic reaches a common interface, whereas in panel (b), $n_{surv}$ denotes the number of simulation runs for which the vulnerable SIS subsystem sustains an active epidemic until the end of the observation interval at $t=10,000$ MCS.
    The active SIR subsystem generates only a transient epidemic wave, whereas the activity induced in the vulnerable SIS subsystem may persist after the SIR infection outbreak has ceased.
    }
    \label{fig:SIR_SIS_density}
\end{figure*}
We next ask whether stabilization of the vulnerable SIS subsystem requires a persistently active neighboring source. 
To address this question, we replace the robust SIS subsystem considered above by an SIR region. 
In addition to changing the source from persistent to transient epidemic activity, this replacement also changes the universality class of the spreading dynamics: 
Whereas the SIS model belongs to the directed percolation (DP) universality class, critical SIR spreading is governed by dynamic isotropic percolation (DIP).
We therefore also aim to determine whether the distinct spatial and temporal structure of infectious clusters generated by SIR dynamics modifies the stabilization mechanism and the correlated population structures induced in the neighboring vulnerable SIS subsystem.
Indeed, the resulting disease spreading dynamics is qualitatively different. 
Since recovered individuals in the SIR subsystem become permanently immune, the infection can spread only once through this (small) region. 
Consequently, the epidemic develops as a single transient invading wave that propagates through the SIR subsystem and eventually leaves behind a population of recovered individuals.

As shown in Fig.~\ref{fig:SIR_SIS_density}, we vary the infection probability in the active SIR subsystem over the range $0.204 \leq \lambda_{\mathrm{a}} \leq 0.216$, while the infection probability in the vulnerable SIS subsystem is held fixed at $\lambda_{\mathrm{v}}=0.109$. 
All values considered for the SIR source region reside above the corresponding epidemic threshold $\lambda_c^{\mathrm{SIR}} \approx 0.203$.
In the active SIR subsystem, the combined density $n_I(t)+n_R(t)$ measures the cumulative extent of the epidemic as the infectious wave propagates through the region. 
Since recovered individuals remain permanently immune, this quantity monotonically increases as the epidemic spreads and eventually approaches a constant value once the transient infectious wave has passed. 
As expected, this final recovered-plus-infected population increases with $\lambda_{\mathrm{a}}$, since larger infection probabilities allow the epidemic to spread through a greater fraction of the SIR subsystem.
The infectious density $n_I(t)$, which directly characterizes the transient active component of the single-peaked epidemic, is graphed in the inset.

In contrast to the SIS source considered above, varying the SIR infection probability over this interval produces only a comparatively modest change in the propagation time of the epidemic wave. 
This behavior is consistent with the system traversal-time results presented in Sec.~\ref{sec:traversal_time}, where the dependence of $t_f$ on the distance from the epidemic threshold was found to be considerably weaker for SIR spreading dynamics than for SIS dynamics over the investigated supercritical range.
Consequently, the SIR epidemic reaches the subsystem interface at rather similar times for the different values of $\lambda_{\mathrm{a}}$ considered here.
This weak dependence of the SIR propagation time is also reflected in the response of the vulnerable SIS subsystem. 
Despite of the variation of $\lambda_{\mathrm{a}}$ in the neighboring SIR region, the onset and subsequent growth of epidemic activity in the vulnerable SIS subsystem occur at similar times for the different curves. 
Thus, unlike for the SIS--SIS case, increasing the infection probability of the source region does not produce a pronounced systematic shift in the time at which the vulnerable subsystem is restabilized. 
Instead, once the transient SIR wave reaches the interface and introduces a sufficiently developed infectious cluster in the vulnerable region, the subsequent evolution is primarily governed by the intrinsic SIS dynamics at the fixed infection probability $\lambda_{\mathrm{v}}=0.109$.

\begin{figure*}[t]
    \centering
    \subfloat[
        Transient epidemic in the active SIR subsystem.
        \label{fig:active_SIR_density_size}
    ]{
        \includegraphics[width=0.47\textwidth]
        {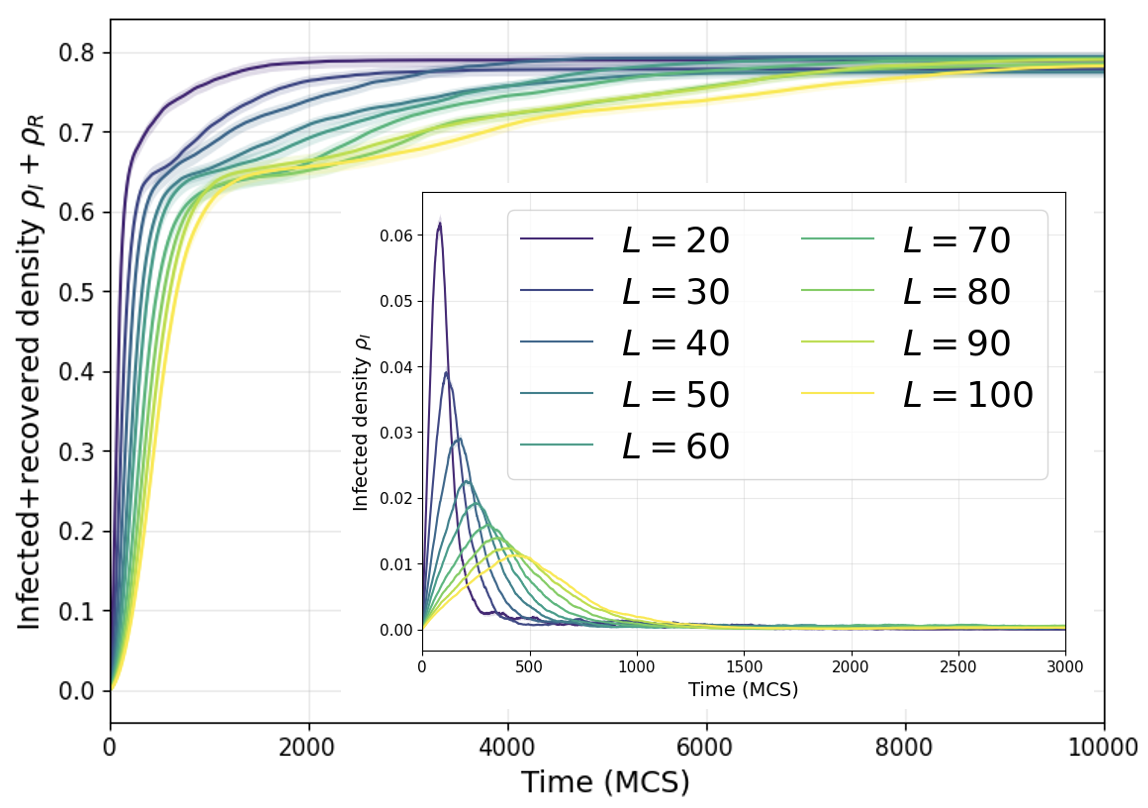}
    }
    \hfill
    \subfloat[
        Vulnerable SIS subsystem.
        \label{fig:vulnerable_SIS_density_size}
    ]{
        \includegraphics[width=0.47\textwidth]
        {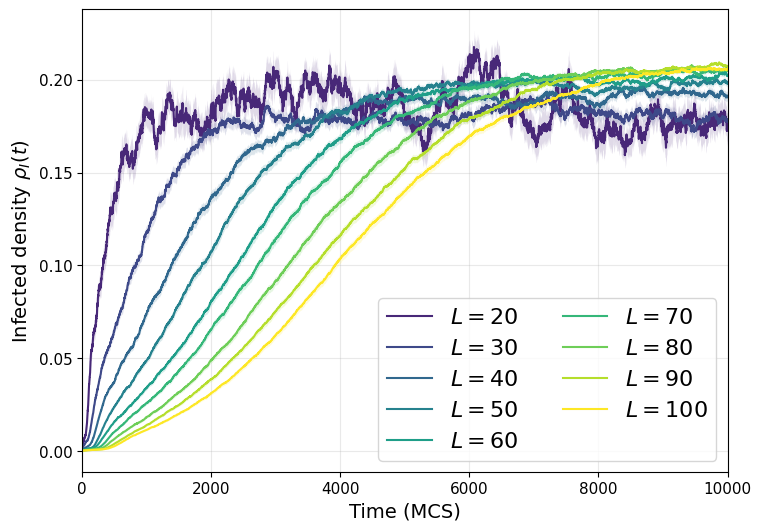}
    }
    \caption{
    Density evolution in the coupled SIR--SIS system for different common linear system sizes $L$.
    The infection probabilities are fixed at   $\lambda_{\mathrm{a}}=0.210$ in the active SIR region and $\lambda_{\mathrm{v}}=0.109$ in the vulnerable SIS region, with recovery probability $\gamma=0.1$ in both subsystems.
    Each region is initialized with a single infected individual; the subsystems remain coupled throughout the entire simulation; and the results are obtained from $1200$ independent stochastic realizations for each system size.
    Panel (a) shows the total affected density $\rho_I^{(A)}(t)+\rho_R^{(A)}(t)$ in the active SIR region ($A$), while its inset displays the transient infectious density $\rho_I^{(A)}(t)$.
    The SIR densities are averaged over realizations in which the epidemic reaches the common interface.
    Panel (b) shows the infected density $\rho_I^{(B)}(t)$ in the vulnerable
    SIS region ($B$), averaged over the corresponding realizations that remain     active until the end of the observation interval at $t=10,000$ MCS.
    Increasing $L$ delays both the propagation of the transient SIR epidemic and the subsequent reactivation of the vulnerable SIS region, while the successfully stabilized SIS systems approach approximately the same quasi-stationary infected density.
    }
    \label{fig:SIR_SIS_density_size}
\end{figure*}
A particularly important observation is provided by the survival counts $n_{surv}$ reported in the figure legends. 
For each value of the SIR infectivity, the number of realizations in which the SIR epidemic successfully reaches the interface here is identical with the number of realizations in which the vulnerable SIS subsystem subsequently remains active. 
Thus, conditional on successful arrival of the transient SIR epidemic at the subsystem boundary, the vulnerable SIS region is invariably re-stabilized in all such realizations within the present ensemble. 
This demonstrates that a persistently active source is not required for epidemic stabilization: 
A single transient SIR infection wave (whose spreading clusters are characterized by he dynamic isotropic percolation universality class) can provide a sufficiently correlated infectious configuration to drive the neighboring extinction-prone SIS subsystem into a long-lived active epidemic state.

\subsubsection{Finite-size dependence of SIR--SIS stabilization}
\label{sec:SIR_SIS_finite_size}

We next investigate how the stabilization in the SIS region induced by a transient SIR epidemic in an adjacent patch depends on the linear size $L$ of the diffusively coupled subsystems. 
The infection probabilities are fixed at $\lambda_{\mathrm{a}}=0.210$ in the active SIR region and $\lambda_{\mathrm{v}}=0.109$ in the vulnerable SIS region, while the recovery probability is set to $\gamma=0.1$ in both subsystems.
The common linear system size is varied over the range $10\leq L\leq100$. 
Each subsystem is initialized with a single infected individual; the two regions remain diffusively coupled throughout the entire observation window until $t=10,000$ MCS.

Figure~\ref{fig:SIR_SIS_density_size} depicts the resulting density evolution for different system sizes. 
In the active SIR subsystem, {\em c.f.} Fig.~\ref{fig:active_SIR_density_size}, the total affected density $\rho_I(t)+\rho_R(t)$ increases as the epidemic spreads and then approaches a constant value after the transient infection wave has passed. 
Since recovered individuals cannot become susceptible again, this long-time value measures the fraction of the SIR region that has been reached by the epidemic, rather than a stationary active density.
The corresponding infectious density $\rho_I(t)$ is displayed in the inset.
Unlike the cumulative affected density, $\rho_I(t)$ exhibits a transient pronounced maximum and subsequently decreases to zero for every system size. 
The peak shifts toward later times and becomes broader as $L$ increases because the epidemic wave requires more time to propagate across a wider region.
Consequently, although the SIR epidemic is necessarily transient, its duration increases with system size. 
The peak infectious density is enhanced in the smallest systems because a localized spreading cluster encompasses a larger fraction of the total population when $L$ is small. 
For bigger systems, the same propagating infection front occupies a smaller fraction of the full lattice at any particular time.
Yet, the final affected density displays only a moderate dependence on system size.
For the smallest systems, a successful outbreak can rapidly cover nearly the entire SIR region.
In larger systems, spatial heterogeneity and the formation of recovered sites behind the advancing front leave a larger fraction of sites unaffected, resulting in a slightly reduced asymptotic value of $\rho_I+\rho_R$. 
Nevertheless, the large final affected densities show that, in the realizations included in the averages, the transient epidemic reaches a substantial fraction of the active subsystem.

The slow late-time increase in the total affected density in Fig.~\ref{fig:active_SIR_density_size} is consistent with transmission from the SIS patch  back into the SIR subsystem. 
After the primary SIR wave reaches the interfaces and reactivates the neighboring SIS region, the two subsystems remain coupled for nearly the entire observation interval. 
In realizations where SIS activity persists, infectious individuals near an interface can therefore infect sites in the SIR region that remained susceptible after the primary wave. 
Even occasional new infections can produce a gradual, uneven increase in $\rho_I^{(A)}(t)+\rho_R^{(A)}(t)$ while the instantaneous infectious density $\rho_I^{(A)}(t)$ remains small.

The response of the vulnerable SIS subsystem is shown in Fig.~\ref{fig:vulnerable_SIS_density_size}. 
The infected density rises earlier in smaller systems because the SIR epidemic has a shorter distance to traverse before reaching the subsystem interfaces.
Increasing $L$ delays the arrival of the SIR wave and therefore shifts the onset of SIS reactivation toward later times. 
Once the invading disease activity has become established, however, the surviving SIS realizations approach approximately the same quasi-stationary infected density $\rho_I^{\mathrm{qs}}\approx0.22$. 
Thus, as in the coupled SIS--SIS system, the system size primarily controls the transient propagation time, whereas the long-time infected density in the successfully stabilized vulnerable region is mainly determined by its fixed infection probability $\lambda_{\mathrm{v}}=0.109$. 
The comparatively large fluctuations for the smallest system sizes naturally reflect the stronger finite-size fluctuations associated with their smaller populations.

The transient nature of SIR dynamics requires a different definition of the source event from that employed for the coupled SIS--SIS system. 
In an SIS region, a realization can be classified according to whether epidemic activity survives until the end of the observation window. 
In contrast, the infectious population in any finite SIR system must eventually vanish because recovered individuals acquire permanent immunity. 
Therefore, SIR survival at long times is neither expected nor an appropriate criterion for a successful source outbreak.
Instead, we classify an SIR realization according to whether the epidemic reaches the subsystem boundary. 
Boundary arrival demonstrates that the infection cluster has spread from its initial location over a distance of order $L$ and has therefore developed on the full scale of the weaker, vulnerable subsystem.
It also identifies the realizations in which the transient SIR epidemic has an opportunity to transmit activity across the interfaces.
We accordingly define the finite-size stabilization probability as
\begin{equation}
P_{\mathrm{stab}}^{\mathrm{SIR}}(L)
= P\!\left(\mathcal{S}_B \mid \mathcal{I}_A\right) \approx \frac{N(\mathcal{S}_B \cap \mathcal{I}_A)}{N(\mathcal{I}_A)} \ ,
\label{eq:SIR_stabilization_probability_size}
\end{equation}
where $A$ denotes the active SIR and $B$ the vulnerable SIS subsystem.
Here, $\mathcal{S}_B$ denotes survival of activity in subsystem $B$, while $\mathcal{I}_A$ denotes the epidemic in subsystem $A$ reaching a coupling boundary.
The denominator $N(\mathcal{I}_A)$ counts the realizations in which this boundary is reached, and the numerator $N(\mathcal{S}_B \cap \mathcal{I}_A)$ tallies those realizations in which activity in subsystem $B$ also survives.
In the simulations, the numerator counts the boundary-reaching SIR realizations for which the vulnerable SIS region remains active until $t=10,000$ MCS, while the denominator counts all SIR realizations that reach the interfaces.

\begin{figure}[t]
    \centering
    \includegraphics[width=0.95\columnwidth]
    {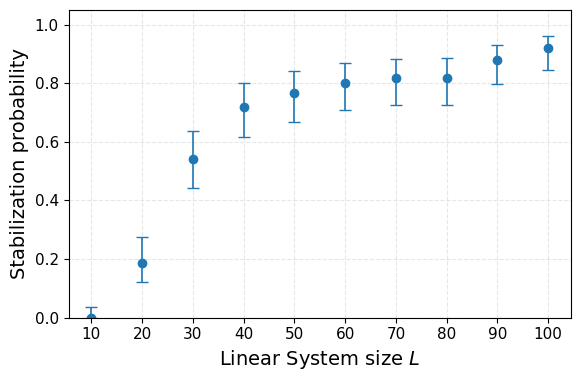}
    \caption{
    Conditional stabilization probability
    $P_{\mathrm{stab}}^{\mathrm{SIR}}(L)$ as a function of the common linear system size $L$.
    The probability is defined as the probability that the vulnerable SIS     subsystem $B$ remains active until $t=10,000$ MCS, conditional on the     transient epidemic in the active SIR subsystem $A$ reaching the common     interface.
    The infection probabilities are fixed at   $\lambda_{\mathrm{a}}=0.210$ in the SIR region and $\lambda_{\mathrm{v}}=0.109$ in the SIS region, with recovery probability $\gamma=0.1$ in both subsystems.
    Each subsystem is initialized with a single infected individual, and each point is obtained from $100$ independent stochastic realizations.
    The conditional probability increases overall with $L$, demonstrating that boundary-reaching SIR outbreaks become increasingly effective at establishing long-lived epidemic activity in the vulnerable SIS region. 
    The subsystems remain coupled throughout the observation interval of $10,000$ MCS.
    }
    \label{fig:SIR_SIS_stabilization_size}
\end{figure}
Figure~\ref{fig:SIR_SIS_stabilization_size} shows that this conditional stabilization probability increases strongly with system size. 
Its growth becomes more gradual for larger systems, increasing from about $0.77$ at $L=50$ to approximately $0.92$ at $L=100$.
The increase of $P_{\mathrm{stab}}^{\mathrm{SIR}}(L)$ demonstrates that reaching the interface is necessary but does not always guarantee successful stabilization, particularly in small systems. 
A boundary-reaching SIR outbreak has a spatial extent of order $L$, but in a small lattice the absolute number of infected sites contained in the propagating cluster and the length of the interface are both limited. 
The activity transmitted into the vulnerable subsystem may consequently be too sparse or too short-lived to develop into a robust SIS population before stochastic fluctuations might manage to eliminate it.

For larger systems, the transient SIR wave may contain a larger absolute number of infected sites and interact with the vulnerable region over a longer interface. 
It can consequently introduce activity at more locations and generate a more spatially extended effective initial condition in the SIS subsystem. 
Once this transferred activity becomes sufficiently established, repeated SIS infection and recovery events may effectively sustain the epidemic outbreak even after the original SIR wave has disappeared. 
The increasing conditional probability therefore supports the conclusion that the stabilization mechanism is controlled not merely by the first arrival of infection at the interface, but also by the spatial extent and population size of the activity transferred into the vulnerable region.
Unlike for the corresponding coupled SIS--SIS result, the conditional probability does however not reach unity over the investigated $L$ range and remains at $\approx 0.92$ for $L=100$.
This marked difference from the SIS--SIS setup is consistent with the transient character of the SIR infection source:
An active SIS region can continuously supply infected individuals while the subsystems remain coupled, whereas the SIR region provides only a finite epidemic pulse. 
Successful stabilization of the weak SIS subsystem must therefore occur within the limited time interval during which the SIR infection front is present at the interface.

Taken together, the density and conditional-probability measurements reveal two {\em complementary} finite-size effects: 
Increasing the system size delays the arrival of the transient SIR epidemic wave and the subsequent reactivation of the vulnerable subsystem, but it also substantially increases the probability that the transferred activity produces lasting stabilization. 
Thus, a shorter propagation time in a small system does not imply a more reliable stabilization process. 
The decisive factor is whether the transient SIR wave can generate a sufficiently extended and populous infectious configuration in the neighboring SIS region before the SIR activity disappears.

\subsubsection{Spatial correlations in coupled SIR--SIS systems}

\begin{figure*}[t]
    \centering
    \subfloat[
        Correlation length parallel to the interface.
        \label{fig:SIR_SIS_xi_parallel}
    ]{
        \includegraphics[width=0.47\textwidth]
        {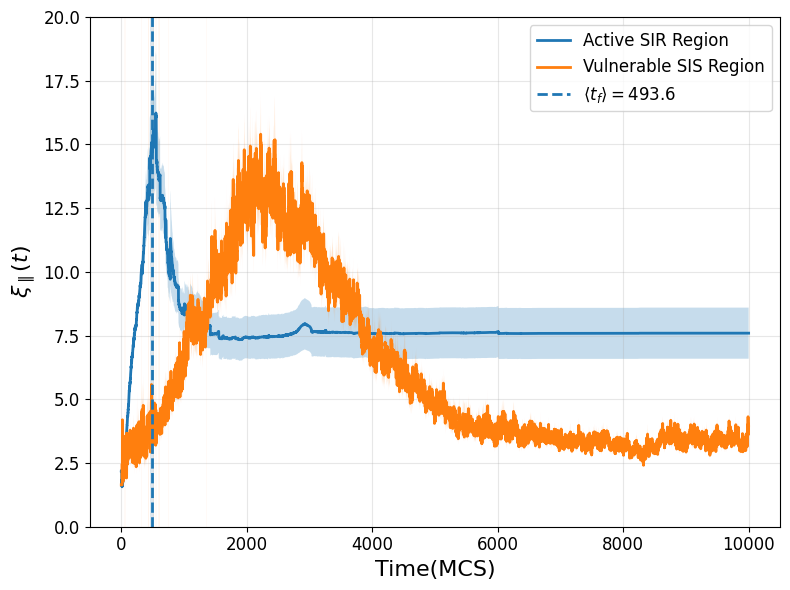}
    }
    \hfill
    \subfloat[
        Correlation length perpendicular to the interface.
        \label{fig:SIR_SIS_xi_perp}
    ]{
        \includegraphics[width=0.47\textwidth]
        {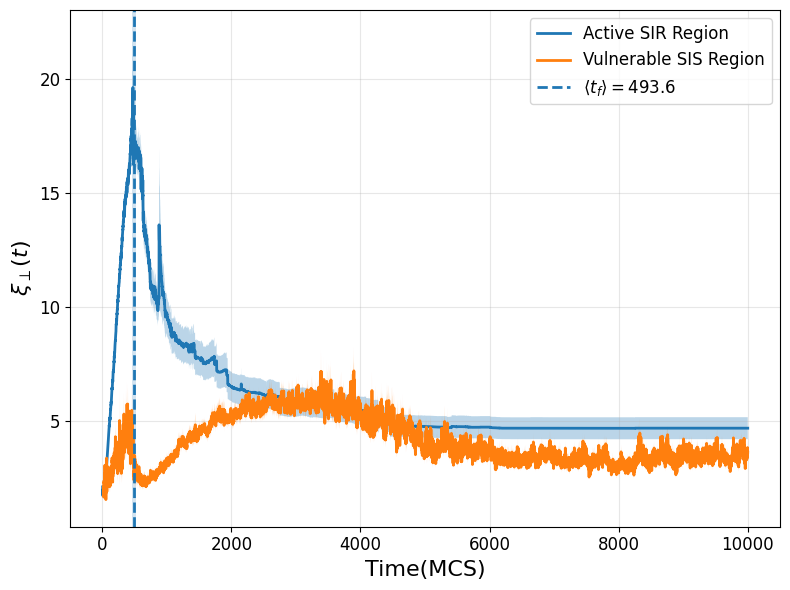}
    }
    \caption{
    Time evolution of the spatial correlation lengths in the active SIR and the vulnerable SIS regions of a coupled SIR--SIS system.
    Both subsystems have linear size $L=100$, with infection probabilities $\lambda_{\mathrm{a}}=0.210$ in the SIR region and $\lambda_{\mathrm{v}}=0.109$ in the vulnerable SIS region; the recovery probability is set to $\gamma=0.1$ in both patches.
    Each subsystem is initialized with a single infected individual; the results are obtained from $100$ independent stochastic realizations.
    Panel (a) shows the correlation length     $\xi_{\parallel}(t)$ parallel to the common interface; panel (b) displays the correlation length $\xi_{\perp}(t)$ perpendicular to the interface.
    The vertical dashed lines mark the mean time $t_f$ at which the transient SIR epidemic first reaches the interface.
    Following this arrival, correlations develop in the vulnerable SIS subsystem, with a stronger and more persistent enhancement parallel to the interface.
    In the active SIR subsystem, the correlation lengths subsequently decrease as the transient infection wave passes and infected individuals enter the permanently recovered state.
    The correlation lengths are extracted from exponential fits to the equal-time spatial correlation functions.
    The subsystems stay coupled for the entire simulation duration ($t_{\mathrm{dec}}=10,000$ MCS.)
    }
    \label{fig:SIR_SIS_xi}
\end{figure*}
We proceed to investigate the emerging spatial structures during the stabilization process when the vulnerable SIS subsystem is coupled to an active SIR region. 
This case is qualitatively different from the SIS--SIS system considered above:
Whereas the active SIS subsystem can sustain epidemic activity indefinitely, the SIR subsystem supports only a transient epidemic wave, since infected individuals eventually recover and cannot become susceptible again.
Nevertheless, as discussed above, this transient spreading process can induce long-lived activity in the neighboring SIS region.

To characterize the spatial organization associated with this process, we analyze the equal-time infection correlation functions in both subsystems.
As for the coupled SIS--SIS system, correlations are measured separately in the directions parallel and perpendicular to the subsystem interface, and the corresponding correlation lengths $\xi_{\parallel}(t)$ and $\xi_{\perp}(t)$ are extracted from the spatial decay of the correlation functions. 
Again, the parallel direction refers to separations along the interface, while the perpendicular direction measures correlations normal to the interface and therefore along the direction in which activity is transmitted from the SIR region into the vulnerable SIS region.

Figure~\ref{fig:SIR_SIS_xi} shows the temporal evolution of these correlation lengths for the active SIR and vulnerable SIS subsystems.
The vertical dashed lines indicate the mean time at which the epidemic activity originating in the SIR region first reaches the interfaces, $t_f = 493.6 \pm 28.9$ MCS.
Before the SIR epidemic arrives at the interfaces, the correlation lengths within the active region increase rapidly as the epidemic cluster expands outward from its initial location.
This growth is particularly pronounced for $\xi_{\perp}(t)$, which reaches its maximum approximately when the spreading epidemic first encounters the subsystem boundary. 
Its rapid increase therefore reflects the spatial expansion of the transient SIR epidemic wave across the active subsystem.
Following the arrival of the epidemic at the interface, the correlation lengths in the SIR subsystem decrease. 
This behavior of course follows from the transient nature of SIR dynamics: 
Sites left behind the propagating infection front enter the recovered state and cannot participate in subsequent infection events. 
Hence, the strongly correlated infectious cluster associated with the advancing epidemic wave is not maintained after the epidemic wave has passed.

The response of the vulnerable SIS subsystem is qualitatively different.
Following the arrival of the SIR epidemic at the interface, spatial correlations begin to develop within the SIS region. 
In the perpendicular direction, $\xi_{\perp}(t)$ increases gradually and reaches a moderate maximum before subsequently decreasing toward its long-time value. 
The effect is considerably more pronounced in the direction parallel to the interface: 
$\xi_{\parallel}(t)$ continues to increase after the SIR front has reached the interface and develops a broad maximum at later times.
The pronounced enhancement of $\xi_{\parallel}$ relative to $\xi_{\perp}$ indicates an anisotropic spatial organization of the activity induced in the vulnerable region. 
Since $\xi_{\parallel}$ measures correlations along the subsystem interface, its increase points to the formation of a spatially extended correlated infection structure along the boundaries between the two regions. 
The smaller perpendicular correlation length, in contrast, reflects the more limited extent of these correlations normal to the interface. 
These features demonstrate the formation of a correlated, approximately planar infection front along the interface, which subsequently induces epidemic activity within the vulnerable SIS subsystem \cite{SimulationMovies}.

An interesting observation in Fig.~\ref{fig:SIR_SIS_xi} is the clear temporal separation between the arrival of the SIR epidemic at the interface and the subsequent maximum of the correlation length in the vulnerable SIS region. 
The SIR wave reaches the interface at approximately $t \simeq 500$ MCS, whereas the strongest correlations parallel to the interface develop considerably later at $\sim 2,200$ MCS. 
The arrival of epidemic activity at the interface therefore marks the onset of the stabilization process. 
Following this initial arrival, substantial additional time is required for the induced activity to spread and develop extended spatial correlations within the vulnerable subsystem.
At later times, the correlation lengths in the SIR subsystem decrease as the transient epidemic wave passes through the active region and infected individuals enter the recovered state. 
In the SIS subsystem, by contrast, finite spatial correlations persist at long times. 
This difference mirrors the distinct dynamics of the two epidemic processes: 
As emphasized before, activity in the SIR region is transient, while infectious seeds introduced into the SIS region can generate continuing activity through repeated infection and recovery events.

These findings fully align with the stabilization mechanism discussed above: 
A transient SIR epidemic wave is capable of inducing spatially correlated epidemic activity along the interface with the vulnerable SIS subsystem. 
The subsequent development of these correlations within the SIS region provides a spatial characterization of the invasion process through which the vulnerable subsystem is driven away from stochastic extinction.

\subsubsection{Comparison with stabilization by a persistent SIS source}

As we have seen, the effect of increasing the infection probability in the more robust source region differs between the SIS--SIS and SIR--SIS systems.
For SIS--SIS coupling, larger values of $\lambda$ generate stronger spreading activity and typically produce invading clusters that reach the vulnerable subsystem at earlier times. 
As a result, the vulnerable region receives incoming epidemic activity sooner, thereby reducing the likelihood of stochastic extinction.
In contrast, the SIR-driven stabilization displays a weaker dependence on the infection probability of the source region. 
Although increasing the infection probability produces somewhat faster propagation of the transient SIR epidemic wave, the subsequent stabilization dynamics of the vulnerable SIS subsystem remains remarkably similar. 
Nevertheless, this difference just reflects the distinct nature of the source dynamics: 
The SIS subsystem can repeatedly generate invading activity, whereas the SIR subsystem produces essentially a single propagating epidemic wave. 
Once that wave reaches the interface and establishes activity in the vulnerable SIS region, the subsequent evolution is governed primarily by the re-incited intrinsic SIS spreading dynamics.

In conclusion, the diffusively coupled SIS--SIS and SIR--SIS model results both demonstrate that large, spatially correlated infectious clusters generated in a neighboring source region can re-excite and stabilize epidemic activity in an extinction-prone SIS subsystem.
Importantly, the source need not remain active indefinitely: 
Even a transient SIR epidemic wave can induce long-lived SIS activity once a sufficiently
developed correlated infectious cluster has entered the vulnerable region.

\begin{figure*}[t]
    \centering
    \subfloat[
        Active SIR subsystem.
        \label{fig:active_SIR_SIR_density_size}
    ]{
        \includegraphics[width=0.47\textwidth]
        {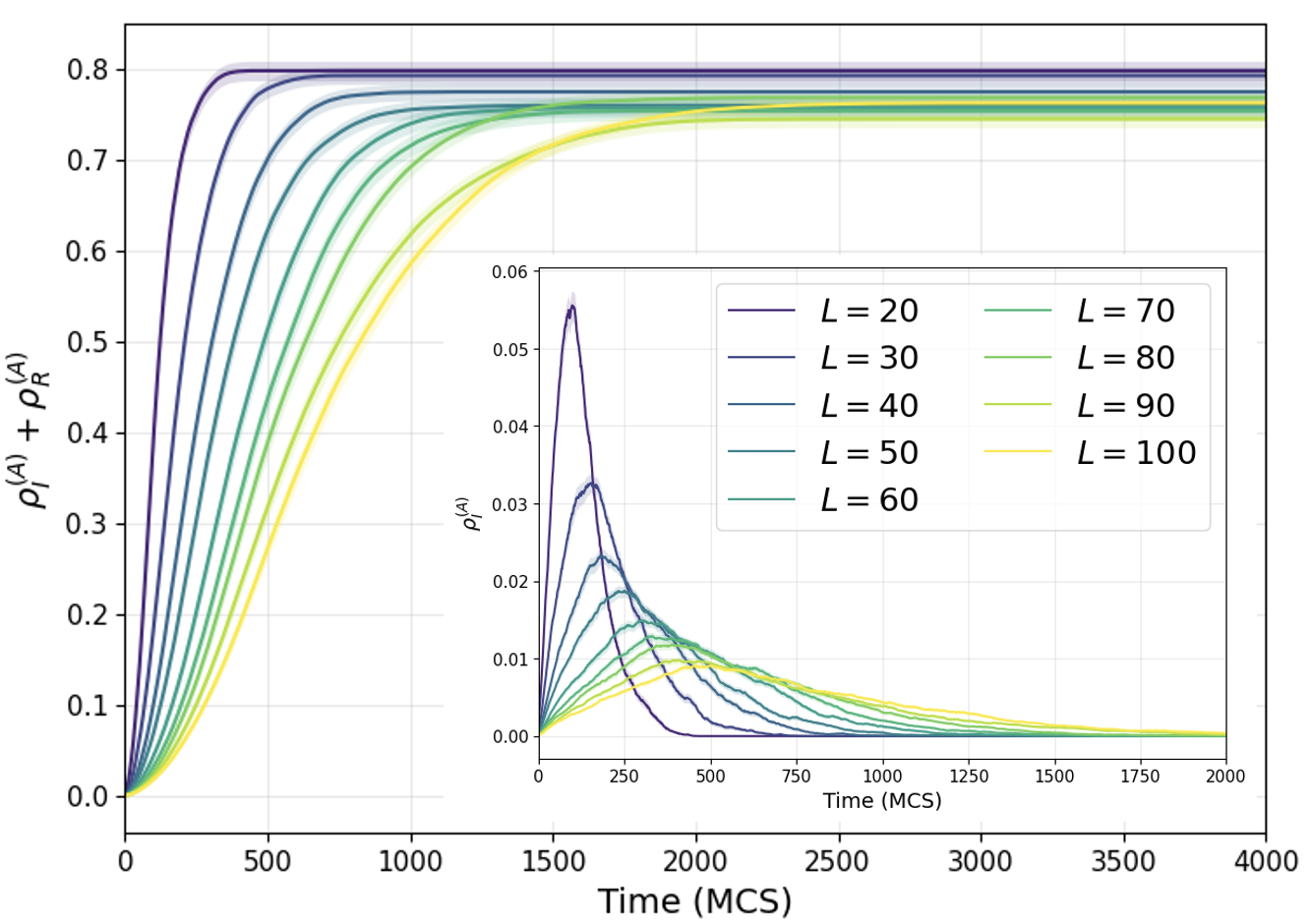}
    }
    \hfill
    \subfloat[
        Vulnerable SIR subsystem.  \label{fig:vulnerable_SIR_SIR_density_size}
    ]{
        \includegraphics[width=0.47\textwidth]
        {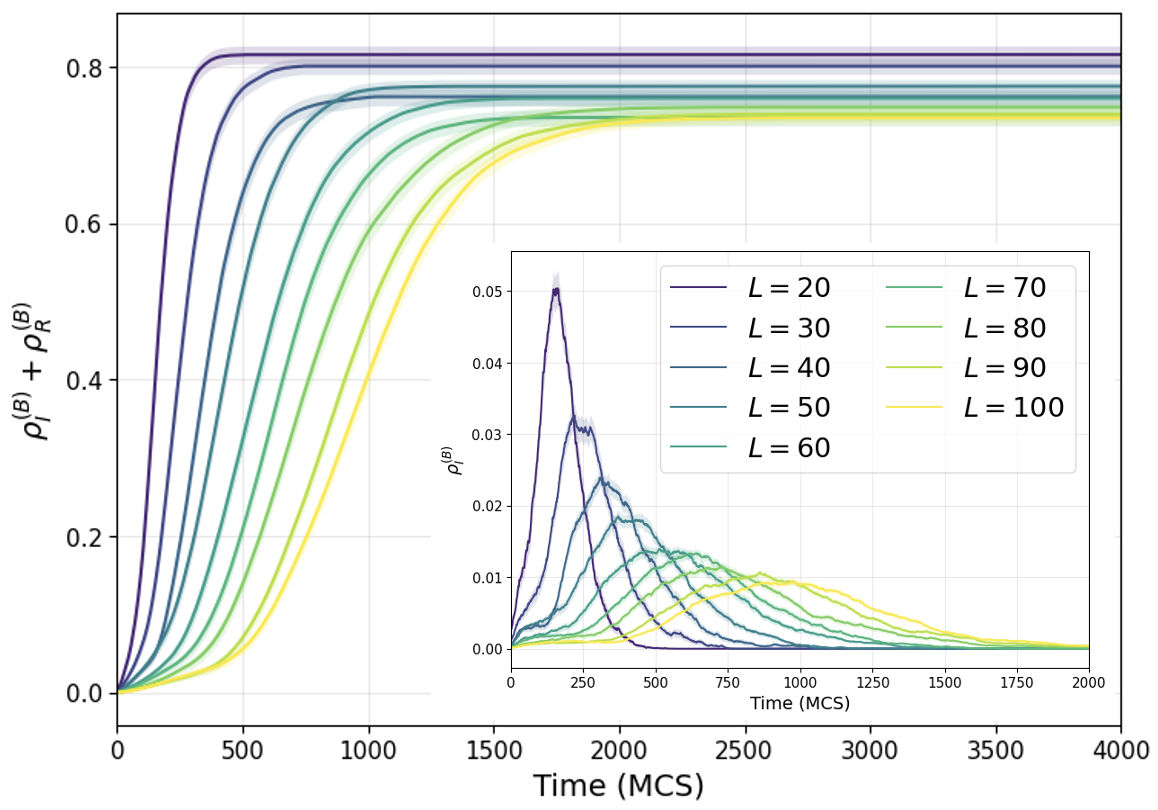}
    }
    \caption{
    Density evolution in two diffusively coupled SIR subsystems for different common linear system sizes $L$.
    The infection probabilities are fixed at   $\lambda_{\mathrm{a}}=0.210$ in the active SIR subsystem and $\lambda_{\mathrm{v}}=0.206$ in the vulnerable SIR subsystem, with common recovery probability $\gamma=0.1$ in both regions.
    Each subsystem is initially seeded with a single infected individual individual at   its center; the subsystems remain coupled throughout the entire observation interval of $10,000$ MCS; and the results are obtained from $100$ independent stochastic realizations for each system size.
    The main graphs show the total affected density $\rho_I(t)+\rho_R(t)$, while the insets display the corresponding transient infectious density $\rho_I(t)$.
    The densities in panel (a) are averaged over realizations in which the epidemic initiated at the center of the active subsystem reaches its boundary.
    The densities in panel (b) are averaged over realizations in which the epidemic transmitted through the interface subsequently reaches the central column of the vulnerable subsystem.
    Solid curves depict the conditional mean densities, and the surrounding shaded regions indicate their standard errors of the mean.
    Increasing $L$ delays both the primary epidemic in the active subsystem and the secondary epidemic in the vulnerable subsystem.
    The infectious densities ultimately vanish in both regions, while the total affected densities approach nonzero final values that decrease with increasing system size.
    }
    \label{fig:SIR_SIR_density_size}
\end{figure*}

\subsection{Transmission between two coupled SIR regions}
\label{sec:SIR_SIR}

We finally consider two diffusively coupled SIR subsystems. 
This configuration differs qualitatively from both the SIS--SIS and SIR--SIS systems discussed above: 
Since recovered individuals acquire permanent immunity in both SIR regions, neither subsystem can support a stationary active state. 
Diffusive coupling therefore cannot stabilize a persistent endemic population in this setup.
Instead, it may allow a transient epidemic wave front originating in the more active SIR subsystem to cross the interface and initiate a secondary macroscopic outbreak in the vulnerable SIR region.
To investigate the finite-size dependence of this transmission process, the common linear system size is varied over the range $10\leq L\leq100$.  
The infection probabilities are fixed at $\lambda_{\mathrm{a}}=0.210$ in the active SIR subsystem and $\lambda_{\mathrm{v}}=0.206$ in the vulnerable SIR subsystem, while the recovery probability is set to $\gamma=0.1$ in both regions. 
Each subsystem is initially seeded with a single infected individual at its center, and the two regions remain coupled for the entire duration of the Monte Carlo simulations (until $t_{\mathrm{dec}}=10,000$ MCS). 
For each system size, the results are obtained from $100$ independent stochastic realizations.

\subsubsection{Finite-size dependence of the epidemic densities}
\label{sec:SIR_SIR_density_size}

Figure~\ref{fig:SIR_SIR_density_size} shows the density evolution in the active and vulnerable SIR subsystems. 
To focus on epidemics that propagate over a macroscopic distance, the densities are averaged over the corresponding successful realizations. 
In the active SIR subsystem, a realization is included if the epidemic initiated at the center reaches the common boundary. 
In the vulnerable SIR subsystem, the density is averaged over realizations in which the activity entering through the interface subsequently reaches the central column of that region. 
The shaded areas represent the standard error of the mean across the selected realizations and therefore quantify the statistical uncertainty of the estimated mean densities.
In each panel, the main graph displays the total affected density $\rho_I(t)+\rho_R(t)$, which measures the cumulative fraction of the population reached by the epidemic. 
The corresponding infectious density $\rho_I(t)$ is plotted in the insets.
Unlike the infected density in the SIS model, the infectious density in either SIR subsystem cannot approach a stationary nonzero value.
Instead, $\rho_I(t)$ exhibits a transient maximum and eventually decreases to zero for every system size. By contrast, the affected density approaches a nonzero plateau because individuals remain in the recovered state permanently after the epidemic has passed.

In the active SIR subsystem, {\em c.f.} Fig.~\ref{fig:active_SIR_SIR_density_size}, increasing the system size shifts the growth of the affected density toward later times. 
The maximum of the infectious density is likewise delayed and becomes broader because the epidemic wave requires more time to propagate across a larger spatial region. 
The peak infectious density is highest for the smallest systems. 
A propagating infection cluster occupies a larger fraction of a small lattice, whereas in a larger system the active epidemic front represents a smaller fraction of the total population at any particular time. After the epidemic wave has traversed the active subsystem, $\rho_I^{(A)}(t)$ vanishes and $\rho_I^{(A)}(t)+\rho_R^{(A)}(t)$ approaches its asymptotic value.
The vulnerable SIR subsystem, {\em c.f.} Fig.~\ref{fig:vulnerable_SIR_SIR_density_size}, displays the same general transient behavior, but the development of the secondary epidemic is delayed relative to the disease outbreak in the active region. 
The infectious activity originating from the single seed in the vulnerable region typically becomes extinct before the invading wave reaches it. 
The subsequent increase of the affected and infectious densities therefore results from reactivation by infected individuals transmitted from subsystem $A$. 
These individuals must first cross the interface and then generate a sufficiently developed infection cluster inside subsystem $B$.

As $L$ increases, the epidemic in subsystem $A$ requires more time to reach the interface, and the transmitted activity must subsequently propagate over a greater distance to reach the center of subsystem $B$. 
The onset of the secondary epidemic and the peak location of $\rho_I^{(B)}(t)$ therefore shift toward later times with increasing system size. 
The delayed response of the vulnerable region directly reflects these two consecutive propagation stages: 
Traversal of the active subsystem followed by invasion of the vulnerable subsystem.
Although the infection and recovery probabilities are identical for all system sizes, the final affected densities display a modest dependence on $L$. 
Each subsystem is initially seeded with a single infected individual, corresponding to an initial infected density $\rho_I(0)=1/L^2$. 
Increasing the system size hence also decreases the initial infected fraction, including that of the primary epidemic in the active SIR region. 
The spatial extent of the system and the initial density can both contribute to finite-size corrections in the fraction of sites ultimately reached by the epidemic. 
This systematic variation is consistent with finite-size effects under the single-seed initialization used here.
The model parameters determine the asymptotic statistical properties of the epidemic, but they do not require the affected density measured in finite, conditionally selected systems to be independent of $L$. 
Recovered sites left behind an advancing SIR infection front cannot be infected again. 
They may consequently block later transmission and isolate susceptible regions that remain unvisited by the epidemic. 
Larger systems provide more space for spatial heterogeneities, branching epidemic clusters, and inaccessible susceptible pockets to develop. 
A boundary- or center-reaching cluster may therefore span a large system while leaving a substantial fraction of the population unaffected. 

\subsubsection{Conditional transmission probability}
\label{sec:SIR_SIR_transmission_probability}

Since the infectious population ultimately vanishes in both SIR subsystems, survival at the end of the observation interval is not an appropriate measure of the coupling outcome. 
We instead characterize the transmission process through the spatial distance reached by the epidemic. 
A realization in subsystem $A$ is classified as a successful source outbreak if the epidemic initiated at its center reaches the common boundary. 
Conditional on this event, we determine whether the activity transmitted into subsystem $B$ propagates from the interface to the center of the vulnerable region.
We denote by $\mathcal{I}_A$ the event that the primary epidemic in subsystem $A$ reaches the common interface, and by $\mathcal{C}_B$ the event that the transmitted epidemic subsequently reaches the central column of subsystem $B$. 
The conditional transmission probability is then defined as
\begin{equation}
P_{\mathrm{trans}}^{\mathrm{SIR-SIR}}(L) = P\!\left(\mathcal{C}_B\mid\mathcal{I}_A\right) \approx \frac{N(\mathcal{C}_B\cap\mathcal{I}_A)}{N(\mathcal{I}_A)} \, .
\label{eq:SIR_SIR_transmission_probability}
\end{equation}
Here, $N(\mathcal{I}_A)$ represents the number of realizations in which the epidemic in subsystem $A$ reaches the interface, while $N(\mathcal{C}_B\cap\mathcal{I}_A)$ denotes the number of these realizations in which the transmitted epidemic subsequently reaches the central column of subsystem $B$.
This conditional probability separates two distinct stages of the transmission process. 
Reaching the interface demonstrates that the primary epidemic has propagated across subsystem $A$, but it does not guarantee that the infections introduced into subsystem $B$ will develop into a macroscopic secondary outbreak. 
The incoming infection cluster may remain localized near the interface and disappear before reaching the interior of the vulnerable region.

\begin{figure}[t]
    \centering
    \includegraphics[width=0.95\columnwidth]
    {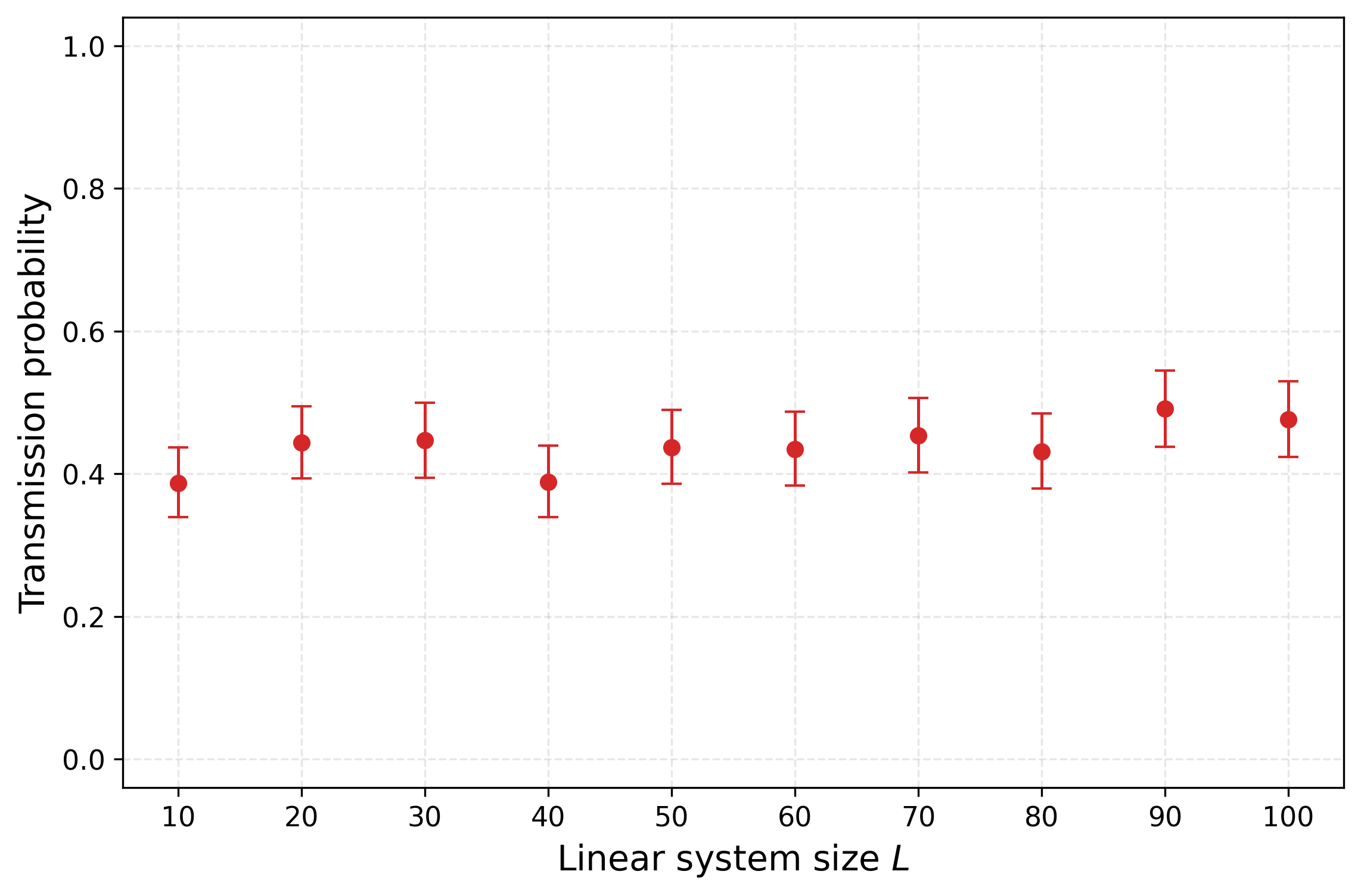}
    \caption{
    Conditional transmission probability     $P_{\mathrm{trans}}^{\mathrm{SIR-SIR}}(L)$ as a function of the common linear system size $L$.
    This quantity is defined as the probability that the epidemic transmitted into the vulnerable SIR subsystem reaches its central column, conditional on the primary epidemic in the active SIR subsystem reaching the common interface.
    The infection probabilities are fixed at   $\lambda_{\mathrm{a}}=0.210$ in the active subsystem and $\lambda_{\mathrm{v}}=0.206$ in the vulnerable subsystem, with common recovery probability $\gamma=0.1$ in both regions.
    Each subsystem is initially seeded with a single infected individual at its center; the subsystems remain coupled throughout the observation interval of $10,000$ MCS;  and each point is obtained from $100$     independent stochastic realizations.
    The conditional transmission probability displays an overall increase with $L$ but remains substantially below $1$ throughout the investigated size range, demonstrating that arrival of the primary epidemic at the interface does not guarantee macroscopic propagation through the vulnerable SIR subsystem.
    }
    \label{fig:SIR_SIR_transmission_size}
\end{figure}
Figure~\ref{fig:SIR_SIR_transmission_size} shows that the conditional transmission probability increases overall from approximately $0.43$ at $L=10$ to approximately $0.61$ at $L=100$. 
The individual values display non-monotonic fluctuations over the investigated size range, as expected for conditional probabilities estimated from a finite ensemble of stochastic realizations. 
For most sizes $L\geq20$, the probability lies between approximately $0.5$ and $0.65$.
The fact that $P_{\mathrm{trans}}^{\mathrm{SIR-SIR}}(L)$ remains substantially below $1$ is physically significant. 
Even when the primary epidemic successfully reaches the interface, the activity transmitted into subsystem $B$ does not necessarily develop into a system-spanning secondary epidemic.
Successful transmission requires not only the arrival of infected individuals at the interface, but also the establishment of a sufficiently large infection cluster that can continue propagating through the vulnerable region before stochastic extinction occurs.

The overall increase of the conditional probability with system size suggests that larger source outbreaks are more effective in spawning an extended epidemic in the neighboring region. 
A boundary-reaching epidemic in a larger active subsystem can encounter the interface over a greater spatial extent and may transmit infected individuals through more interface sites. 
The resulting infection pattern constitutes a larger and more spatially distributed effective initial condition in subsystem $B$, which reduces the probability of immediate extinction following transmission.
Nevertheless, the increasing trend is considerably weaker than the finite-size enhancement of stabilization observed when the receiving subsystem follows SIS dynamics. 
In the SIR--SIS system, repeated susceptible--infected--susceptible transitions allow the activity introduced into the vulnerable region to develop into a long-lived active population. 
In the present SIR--SIR system, recovered sites remain immune whence the secondary epidemic can propagate through the population only once. 
Consequently, both the source and the induced epidemic remain transient phenomena, and successful boundary arrival in subsystem $A$ guarantees neither establishment of disease activity nor its traversal through subsystem $B$.

These results complete the comparison among the three coupled-system configurations. 
In the SIS--SIS system, persistent source activity may repeatedly seed the vulnerable region and ultimately establish a long-lived active state. 
In the SIR--SIS system, even a single transient source wave can generate persistent activity because the receiving SIS subsystem permits reinfection. 
In the SIR--SIR system, by contrast, the source and induced epidemics are both transient. 
Diffusive coupling can transmit the infection and generate a macroscopic secondary outbreak, but it cannot produce a stationary active disease transmission state when permanent immunity is present in both subsystems.

\subsection{Competition between epidemic activity propagation and establishment time scales}
\label{sec:timescale_competition}

The preceding findings identify the propagation and subsequent establishment of incoming epidemic activity as the central mechanism governing the reactivation and stabilization of the vulnerable subsystem that is prone to stochastic extinction. 
In the simulations executed here, both subsystems are initially seeded with a single infected individual. 
Owing to its small size and proximity to the epidemic threshold, the initial activity in the vulnerable SIS subsystem typically becomes extinct before the infection front generated in the more robust source region reaches the interface. 
The subsequent increase of the infected population in the vulnerable region therefore does not represent the continued growth of its original infection cluster. 
Instead, it constitutes a genuine reactivation caused by activity transmitted from the neighboring source subsystem.

This observation modifies the usual interpretation of the competition between propagation and extinction time scales. 
The incoming epidemic front does not need to reach the vulnerable subsystem before its initial activity becomes extinct. 
Although the disease-free configuration is absorbing for an isolated SIS system, it is effectively no longer in an absorbing state when the subsystem is coupled to an active neighboring region, since infected individuals can enter across the common interfaces. 
Consequently, even after the vulnerable region has reached the disease-free state, the invading front may introduce a new, spatially correlated infectious population cluster and re-ignite the local SIS dynamics.
The relevant {\em three time scales} are therefore the system traversal time $t_f$, the decoupling time $t_{\mathrm{dec}}$, and the additional establishment time $t_{\mathrm{est}}$ required after the first boundary arrival. 
Reactivation requires the coupling to persist at least until the source activity reaches the interface,
\begin{equation}
t_{\mathrm{dec}}>t_f \, .
\label{eq:reactivation_condition}
\end{equation}
Reliable autonomous stabilization after decoupling however imposes the stronger condition
\begin{equation}
t_{\mathrm{dec}}\gtrsim t_f+t_{\mathrm{est}} \, .
\label{eq:stabilization_condition}
\end{equation}
Here, $t_{\mathrm{est}}$ denotes the collective time required for the transmitted activity to penetrate the vulnerable subsystem and develop into a {\em sufficiently} extended correlated active population. 
Boundary arrival may initiate reactivation, but it does not by itself guarantee that the induced activity will remain viable after the two regions are separated.

For the coupled SIS--SIS system with $L=100$, epidemic activity reaches the interface after the mean traversal time $t_f=1769.1\pm40.4$ MCS. 
The stabilization probability nevertheless becomes close to unity only if the subsystems remain coupled until approximately $t_{\mathrm{dec}}\gtrsim2700$ MCS. 
The difference between these times, $t_{\mathrm{dec}} - t_f \approx 900$ MCS, provides an estimate of the additional time interval required for the incoming activity to spread within the vulnerable region and establish a robust correlated infectious seed. 
The concurrent increase of the correlation length parallel to the interface supports this interpretation.

Our finite-size results further demonstrate that a short traversal time does not necessarily imply robust stabilization. 
The invading epidemic reaches the interface earlier in smaller systems because the propagation distance is shorter. 
Nevertheless, the stabilization probability remains low because the transferred infection cluster contains comparatively few infected sites, its spatial extent is quite limited, and demographic fluctuations may readily eliminate the newly introduced disease activity. 
As the system size is increased, the source generates larger system-spanning infection clusters and transfers activity across a longer interface. 
The reactivated population can then develop more extended spatial correlations, rendering its subsequent persistence increasingly likely.

The same general mechanism operates when the activity source follows SIR rather than SIS dynamics.
In this case, the source region provides merely a finite interval of infectious activity because the SIR epidemic ultimately disappears after leaving behind recovered individuals. 
Nevertheless, a boundary-reaching SIR wave can reactivate the disease-free SIS subsystem through implanting a new extended correlated infectious cluster across and beyond the interface. 
If this transferred activity becomes sufficiently established before the transient SIR wave disappears, the intrinsic SIS dynamics may subsequently sustain a long-lived active epidemic state. 
Thus, neither survival of the original infection in the vulnerable region nor a permanently active source is required. 
The decisive stabilization condition is that the coupling remain available long enough for the source-generated activity to arrive and establish a sufficiently extended correlated population within the vulnerable subsystem.

\section{Discussion and conclusion}

The results presented in this work demonstrate how spatial, here diffusive, coupling can qualitatively alter stochastic extinction in finite heterogeneous epidemic systems. 
In any finite isolated SIS system, demographic fluctuations will eventually drive the population into the absorbing disease-free state, even when the corresponding infinite system supports an active phase. 
For small systems close to the epidemic threshold, large relative fluctuations may cause this extinction to occur during the earliest stages of the dynamics, before the initial infection cluster has developed into a sufficiently extended and robust active population. 
Coupling the vulnerable subsystem to a neighboring region that supports more robust epidemic activity can countermand this early, stochastically driven extinction. 
Once the infection front generated in the source region reaches the interface, it may introduce a spatially extended and correlated infectious population into the vulnerable region. 
The invasion process therefore effectively replaces the unfavorable single-seed initial configuration with a more disease-favorable {\em correlated} initial state, thereby reactivating the vulnerable subsystem and substantially reducing its probability of subsequently undergoing stochastic extinction \cite{DistefanoShabaniTauber2025}.

The isolated-system results reported in Sec.~III provide the dynamical context for this stabilization mechanism. 
Although every finite SIS system ultimately reaches the absorbing disease-free state, its extinction dynamics close to the epidemic threshold still reflects the universal critical behavior of the corresponding infinite system. 
In particular, we observe approximate power-law behavior in the extinction time as a function of both the control parameter distance from the epidemic threshold and the finite system size. 
Thus, even for the comparatively small SIS lattices investigated here, approaching the critical region produces pronounced critical slowing-down and a strong growth of the characteristic spatial and temporal scales. 
The associated scaling behavior is consistent with the general framework of absorbing-state phase transitions and the directed percolation universality class \cite{Hinrichsen2000,Odor2004,Tauber2014}. 
The exponents extracted from the present simulations should nevertheless be interpreted as effective crossover values, since finite-size, finite-time, and off-critical corrections remain significant. 
The pertinent findings here are not precise numerical values of individual fitted DP scaling exponents, but the observation that the disease extinction and propagation dynamics exhibit clear algebraic behavior as the finite system approaches the critical regime.

The comparison among the three coupled spatially inhomogeneous model configurations shows that the long-time outcome is primarily determined by the dynamics of the receiving weaker region. 
In the diffusively coupled SIS--SIS system, the persistent source repeatedly seeds the vulnerable region and may thereby establish a long-lived active disease-spreading state.
In the coupled SIR--SIS system, even a single transient epidemic wave can generate persistent activity because the target SIS region permits reinfection. 
An indefinitely active source is therefore not required for activity stabilization in the extinction-prone subsystem, provided that the transient invasion front is sufficiently developed when it enters the vulnerable region.
In the SIR--SIR system, by contrast, permanent immunity prevents either subsystem from sustaining stationary activity. 
Coupling may initiate a macroscopic secondary outbreak, but both epidemics ultimately remain transient \cite{Grassberger1983,CardyGrassberger1985,TomeZiff2010, MukhamadiarovTauber2022}.
The decoupling results further demonstrate that the first arrival of infected individuals at the interface does not by itself guarantee autonomous stabilization. 
The two subsystems must remain in diffusive contact not only until the source epidemic reaches the interface, but also for a sufficiently long time afterward to allow the transmitted activity to penetrate the vulnerable region and generate an {\em extended}, spatially correlated infectious population nucleus.
Consequently, the decoupling time must exceed the traversal time, $t_{\mathrm{dec}}>t_f$, by a sufficiently large time interval for this correlated infection seed to become sufficiently robust to persist autonomously after the coupling is removed.

The correlation-length measurements provide a quantitative spatial characterization of the disease revival and establishment process. 
After the invasion front reaches the vulnerable subsystem, correlations grow particularly strongly in the direction parallel to the interface, indicating the formation of an extended, approximately planar infectious structure along the boundary. 
The continued development of these correlations after the first boundary arrival confirms that stabilization is a collective spatio-temporal process rather than the consequence of a single successful transmission event. 
This also explains why smaller systems may exhibit shorter traversal times but lower stabilization probabilities: 
Activity reaches the interface quickly, yet the transmitted cluster remains small and susceptible to detrimental demographic fluctuations. 
Increasing the system size consequently produces competing effects. 
The source front must travel a greater distance before reaching the interface, but a larger source region can also generate a more extended boundary-reaching cluster and transfer activity across longer interfaces. 
The amount and spatial organization of the transmitted activity may therefore be more important for successful stabilization than the first-arrival time.

More broadly, our novel findings demonstrate that stochastic extinction in a finite population cannot always be inferred from its local control parameters in isolation. 
In a spatially heterogeneous environment, the fate of a vulnerable region also depends on the dynamics of neighboring populations, the duration (and nature) of the regions' coupling, and the spatial structure of the incoming activity seed. 
Transient coupling may be sufficient to establish long-lived epidemic activity even after the source becomes inactive or the connection is removed. 
For epidemic control, this implies that local disease elimination may be countermanded if neighboring regions are capable of later inducing correlated epidemic invasion fronts or clusters. 
Future studies should examine how this mechanism is modified in more realistic complex model settings that incorporate, {\em e.g.}, unequal subsystem sizes, irregular interface shapes, long-range mobility, temporal variations in the coupling, and spatial disorder. \\

\acknowledgments
We thank Adewale Arojowolu, Henry Field Crumley, Kenny Distefano, and Michel Pleimling for valuable discussions and helpful input.
Research was sponsored by the Army Research Office and was accomplished under Grant Number W911NF-26-1-A323. 
The views and conclusions contained in this document are those of the authors and should not be interpreted as representing the official policies, either expressed or implied, of the Army Research Office or the U.S. Government. 
The U.S. Government is authorized to reproduce and distribute reprints for Government purposes notwithstanding any copyright notation herein.

\end{document}